\documentclass[preprint,12pt]{elsarticle}

  \usepackage[usenames]{color}
  \definecolor{dark-green}{RGB}{0,100,0}
  \usepackage{amsmath}
  \usepackage{bm}
  \newcommand{\Espaco}{\rule[-5mm]{0mm}{13mm}} 
  \newcommand{\de}[1]{\left(#1\right)}

\usepackage{graphics}
\usepackage{graphicx}
\usepackage{epsfig}
\usepackage{amssymb}

\journal{Chaos, Solitons and Fractals }

\begin{document}

\begin{frontmatter}

\title{{\bf Inter-occurrence times and universal laws in finance, earthquakes and genomes$^*$} \footnote{$^*$Invited review to appear in Chaos, Solitons and Fractals.}}

\author{Constantino Tsallis}
\ead{tsallis@cbpf.br}

\address{ Centro Brasileiro de Pesquisas Fisicas, and National Institute of Science and Technology for Complex Systems - Rua Xavier Sigaud 150, 22290-180 Rio de Janeiro-RJ, Brazil\\
and\\
Santa Fe Institute - 1399 Hyde Park Road, Santa Fe, NM 87501, USA}

\begin{abstract}
A plethora of natural, artificial and social systems exist which do not belong to the Boltzmann-Gibbs (BG) statistical-mechanical world, based on the standard additive entropy $S_{BG}$ and its associated exponential BG factor. Frequent behaviors in such complex systems have been shown to be closely related to $q$-statistics instead, based on the nonadditive entropy $S_q$ (with $S_1=S_{BG}$), and its associated $q$-exponential factor which generalizes the usual BG one. In fact, a wide range of phenomena of quite different nature exist which can be described and, in the simplest cases, understood through analytic (and explicit) functions and probability distributions which exhibit some universal features. Universality classes are concomitantly observed which can be characterized through indices such as $q$. We will exhibit here some such cases, namely concerning the distribution of inter-occurrence (or inter-event) times in the areas of finance, earthquakes and genomes. 
\end{abstract}

\begin{keyword}
Complex systems \sep Nonextensive statistical mechanics \sep Nonadditive entropies \sep Finances \sep Earthquakes \sep Genomes
\end{keyword}
\end{frontmatter}

\section{HISTORICAL AND PHYSICAL MOTIVATIONS}

In 1865 Clausius introduced in thermodynamics, and named, the concept of {\it entropy} (noted $S$, probably in honor of Sadi Carnot, whom Clausius admired) \cite{Clausius1865}. It was introduced on completely macroscopic terms, with no reference at all to the microscopic world, whose existence was under strong debate at his time, and still even so several decades later. One of the central properties of this concept was to be thermodynamically {\it extensive}, i.e., to be proportional to the size of the system (characterized by its total mass, for instance).
In the 1870's Boltzmann \cite{Boltzmann1872,Boltzmann1877} made the genius connection of the thermodynamical entropy to the microcosmos.  This connection was refined by Gibbs a few years later \cite{Gibbs1902}. From this viewpoint, the thermodynamic extensivity became the nowadays well known property that the total entropy of a system should be proportional to $N$, the total number of its microscopic elements (or, equivalently, proportional to the total number of microscopic degrees of freedom).  More precisely, in the $N \to\infty$ limit, it should asymptotically be
\begin{equation}
S(N) \propto N \,,
\label{extensive1}
\end{equation}
hence
\begin{equation}
0< \lim_{N \to\infty} \frac{S(N)}{N} < \infty  \,.
\label{extensive2}
\end{equation}
For a $d$-dimensional system, $N \propto L^d$, where $L$ is a characteristic linear size and $d$ is either a positive integer number (the standard dimension, basically), or a positive real number (fractal dimension, a concept which was in fact carefully introduced by Hausdorff and fruitfully explored by Mandelbrot). Consequently, Eqs. (\ref{extensive1}) and (\ref{extensive2}) can be rewritten as follows:   
\begin{equation}
S(L) \propto L^d \,,
\label{extensive3}
\end{equation}
hence
\begin{equation}
0< \lim_{L \to\infty} \frac{S(L)}{L^d} < \infty  \,.
\label{extensive4}
\end{equation}
The entropic functional introduced by Boltzmann and Gibbs (and later on adapted to quantum and information-theoretical scenarios by von Neumann and Shannon respectively) is given (for systems described through discrete random variables) by  
\begin{equation}
S_{BG}(N) = -k \sum_{i=1}^{W(N)} p_i \ln p_i  \,\,\,\, \Bigl(\sum_{i=1}^{W(N)} p_i =1 \Bigr) \,,
\label{BG1}
\end{equation}
where $k$ is conventional positive constant (usually taken to be Boltzmann constant $k_B$ in physics, and $k=1$ in several other contexts), and $i$ runs over all non-vanishing-probability microscopic configurations of the $N$-sized system, $\{p_i\}$ being the corresponding probabilities. In the particular case of equal probabilities, i.e., $p_i=1/W(N) \, (\forall i)$, we recover the celebrated Boltzmann formula
\begin{equation}
S_{BG}(N) = k \ln W(N) \,.
\label{BG2}
\end{equation}
It is clear that, if the microscopic random variables are probabilistically (strictly or nearly) {\it independent}, we have
\begin{equation}
W(N) \propto \mu^N \,\,\,(\mu >1; \, N \to\infty) \,,
\label{BG3}
\end{equation}
hence Eq. (\ref{BG2}) implies that $S_{BG}(N) \propto N$, thus satisfying the (Clausius) thermodynamic expectation of extensivity, here represented by Eq. (\ref{extensive1}). If we have $N$ coins (dices), then $\mu =2$ ($\mu =6$); if we have a $d$-dimensional first-neighbor-interacting Ising ferromagnet in thermal equilibrium with a thermostat,
then $\mu$ essentially is some temperature-dependent real number. 

$W(N)$ might however have a functional dependance  drastically different from (\ref{BG3}). For example, it could be (see \cite{Tsallis1988,GellMannTsallis2004,Tsallis2004,TsallisGellMannSato2005a}, and pages 66-68 of \cite{Tsallis2009}; see also \cite{Tsallis2004Cagliari,Tsallis2009b})
\begin{equation}
W(N) \propto N^\rho \,\,\,(\rho >0; \, N \to\infty) \,,
\label{rhocorrelation}
\end{equation}
or (see page 69 of \cite{Tsallis2009})
\begin{equation}
W(N) \propto \nu^{N^\gamma} \,\,\,(\nu >1; \, 0< \gamma <1; \, N \to\infty) \,.
\label{gammacorrelation}
\end{equation}
Such cases\footnote{Usually $W$ increases with $N$, but it is not forbidden that it asymptotically decreases with $N$ (see \cite{CuradoTempestaTsallis2016} for instance). Therefore, it might in principle also occur $\mu<1$ in Eq. (\ref{BG3}), $\rho<0$ in Eq. (\ref{rhocorrelation}), and $\nu <1$ with $\gamma>0$ in Eq. (\ref{gammacorrelation}). Of course, in such pathological cases, an additive constant of the order of unity must be included in the asymptotic behaviors in order to never violate $W \ge1$. 
} clearly correspond to probabilistically {\it strong} correlations, of different nature though. We easily verify that, for $N\to\infty$,
\begin{equation}
1 << N^\rho << \nu^{N^\gamma} << \mu^N \,.
\end{equation}
This is directly related to strong restrictions which mandate an occupancy of the entire phase space {\it substantially lesser than full (or nearly full) occupancy}  (which corresponds in turn to Eq. (\ref{BG3}), and, for nonlinear dynamical systems, to ergodicity). We may say alternatively that Eq. (\ref{BG3}) is to be associated with an occupancy of phase space with {\it finite} Lebesgue measure, whereas Eqs. (\ref{rhocorrelation}) and (\ref{gammacorrelation}) typically correspond to an occupancy with {\it zero} Lebesgue measure\footnote{Let us further analyze this case. If we have $N$ distinguishable particles, each of them living in a continuous $D$-dimensional space ($D=2d$ if the system is defined in terms of canonically conjugate dynamical variables of a $d$-dimensional system; for example, Gibbs $\Gamma$ phase space is a $2dN$-dimensional space), then the full space of possibilities is a $DN$-dimensional hypercube whose hypervolume equals $D^N$ (under the assumption that virtually all these possibilities have nonzero probability to occur). In such a case, its Lebesgue measure scales precisely as $W(N) \sim D^N$, in conformity with Eq. (\ref{BG3}) with $\mu=D$. Strong correlations in such a system {\it can not increase} its Lebesgue measure, but can of course decrease it, and even make it to be {\it zero}, as are the cases corresponding to Eqs. (\ref{rhocorrelation}) and (\ref{gammacorrelation}). However, in remarkable contrast with the standard situation represented by Eq. (\ref{BG3}), systems do exist whose total number of possibilities can increase even faster than $\mu^N$. Such is the case of $N$ ranked elements. Indeed, the amount of all possible rankings yields $W(N) = N!$ \cite{Herrmann2013}. Consequently the BG entropy given by Eq. (\ref{BG1}) yields $S_{BG}(N) \sim k N \ln N$, which does not conform to thermodynamics. What precise entropic form would recover extensivity for such a case is at present an interesting open question.}.

If we assume --- and we do, for reasons to be presented hereafter --- that entropic extensivity (i.e., Eq. (\ref{extensive1})) must hold {\it in all cases}, we are forced to generically abandon  the BG functional (\ref{BG1}) whenever probabilistically strong correlations are generically present in the system.  This is the primary physical and mathematical origin of the nonadditive entropies introduced in \cite{Tsallis1988} in order to generalize the BG entropy $S_{BG}$ and also concomitantly generalize the BG statistical mechanics. This is fully consistent with crucial remarks by Boltzmann, Gibbs, Fermi, Majorana, Tisza, Landsberg, and various others (see, for instance, Chapter 1 of \cite{Tsallis2009}) pointing the limits of validity of the BG basic hypothesis. In the next Section we show how {\it nonadditive} entropic functionals (e.g., $S_q$ introduced in \cite{Tsallis1988} in order to generalize the BG theory) become mandatory in order to satisfy this demand in those cases which overcome the usual BG frame and its {\it additive} functional $S_{BG}$.

\section{THERMODYNAMICAL ENTROPIC EXTENSIVITY IN \\
STRONGLY CORRELATED SYSTEMS GENERICALLY \\
MANDATES NONADDITIVE ENTROPIC FUNCTIONALS}

In what follows we shall refer to {\it uncorrelated} or {\it weakly correlated} $N$-body systems whenever Eq. (\ref{BG3}) occurs, and to {\it strongly correlated} ones whenever zero-Lebesgue-measure behaviors such as those in Eqs. (\ref{rhocorrelation}) and (\ref{gammacorrelation}) occur. 

Let us introduce now the following entropic functional ($q \in \cal{R}$):
\begin{equation}
S_q = k\frac{1-\sum_{i=1}^W p_i^q}{q-1} \;\;\;(S_1=S_{BG}) \,.
\label{sq}
\end{equation}
This expression can be equivalently rewritten as follows:
 \begin{equation}
S_q = k \sum_{i=1}^W p_i \ln_q \frac{1}{p_i}=-k \sum_{i=1}^W p_i^q \ln_q p_i=-k \sum_{i=1}^W p_i \ln_{2-q} p_i \,,
\end{equation}
where
\begin{equation}
\ln_qz \equiv \frac{z^{1-q}-1}{1-q} \;\;\; (\ln_1 z=\ln z)\,.
\end{equation}
For the particular instance of equal probabilities (i.e., $p_i=1/W$) we have
\begin{equation}
S_q = k \ln_q W=k\frac{W^{1-q}-1}{1-q} \,.
\label{sqequal}
\end{equation}
Consequently, in the case corresponding to Eq. (\ref{rhocorrelation}), we do not wish to use the BG entropy. Indeed, it yields $S_{BG}(N) \propto \ln N$, which violates thermodynamic extensivity. If we use instead Eq. (\ref{sqequal}) we obtain
\begin{equation}
S_{1-1/\rho}(N) \propto N \, ,
\end{equation}
which is thermodynamically admissible!
We can straightforwardly verify that the entropic functional $S_q$ is {\it nonadditive} (in contrast with the {\it additive} functional $S_{BG}$ \cite{Penrose1970}). Indeed, if $p_{ij}^{A+B}=p_i^A p_j^B$, we have
\begin{equation}
\frac{S_q(A+B)}{k}=\frac{S_q(A)}{k}+\frac{S_q(B)}{k}+(1-q)\frac{S_q(A)}{k}\frac{S_q(B)}{k} \,.
\end{equation}
Let us consider now the case corresponding to Eq. (\ref{gammacorrelation}), there is no value of $q$ that would make $S_q(N)$ to be extensive. We are therefore obliged to postulate another entropic functional. Let us define ($\delta \in \cal{R}$) \cite{Tsallis2009,Ubriaco2009}
\begin{equation}
S_\delta=k\sum_{i=1}^W p_i \Bigl[\ln \frac{1}{p_i} \Bigr]^\delta \;\;\; (S_1=S_{BG}) \,.
\end{equation}
If we have equal probabilities, we verify
\begin{equation}
S_\delta=k \Bigl[\ln W\Bigr]^\delta \,.
\end{equation}
We can therefore check that, for Eq. (\ref{gammacorrelation}), 
\begin{equation}
S_{1/\gamma}(N) \propto N \, ,
\end{equation}
which, once again, is thermodynamically admissible. Therefore, once again, in order to achieve thermodynamical extensivity we are led to use a nonadditive entropy. Indeed, if the probabilistic systems $A$ and $B$ are independent (i.e., $p_{ij}^{A+B}=p_i^A p_j^B$), we verify that generically $S_\delta(A+B) \ne S_\delta(A) + S_\delta(B) $. 

In fact, $S_{BG}$, $S_q$ and $S_\delta$ can be unified through \cite{TsallisCirto2013}
\begin{equation}
S_{q,\delta} = k \sum_{i=1}^W p_i \Big[\ln_q \frac{1}{p_i}\Bigr]^\delta \,.
\end{equation}
We verify that $S_{1,1}=S_{BG}$, $S_{q,1}=S_q$ and $S_{1,\delta}=S_{\delta}$. The statistical mechanics associated with $S_\delta$ and $S_{q,\delta}$, as well as the corresponding  nonlinear Fokker-Planck, have been worked out in \cite{RibeiroTsallisNobre2013,RibeiroNobreTsallis2014}. Also, for the equal-probability cases, a generic thermodynamical discussion can be found in \cite{HanelThurner2011a,HanelThurner2011b,HanelThurnerGellMann2014}. Many other entropic functionals can be found in \cite{Tempesta2011,Tempesta2015}; in particular, an intriguing connection between $S_q$ and the Riemann zeta function is exhibited in \cite{Tempesta2011}. The crucial distinction between entropic additivity and entropic extensivity is illustrated in Table  \ref{tableBG}.
\begin{table}
\vspace{-2.0cm}
\begin{tabular}{|c|c|c|c|}   \cline{2-4} 
\multicolumn{1}{c|}{}                                &\multicolumn{3}{c|}{\rule[-3mm]{0mm}{9mm}\footnotesize{\textbf{ENTROPY}} } \\  \hline
\rule[-1mm]{0mm}{7mm}$\bm{W\de{N}}$                  &                   $S_{BG}$                                 & $S_{q}$                                                       & $S_{\delta}$                                                 \\
\rule[-2mm]{0mm}{7mm}$\bm{\de{N \to \infty}}$        &                                                            & $\de{q\neq1}$                                                 & $\de{\delta\neq1}$                                           \\
\rule[-3mm]{0mm}{7mm}                                & \textcolor{dark-green}{\textbf{\footnotesize{(ADDITIVE)}}} & \textbf{\textcolor{dark-green}{\footnotesize{(NONADDITIVE)}}} &\textbf{\textcolor{dark-green}{\footnotesize{(NONADDITIVE)}}} \\ \hline 
\Espaco$\displaystyle{{e.g.,\mu^{N}} \atop \de{\mu > 1} }$  &\textbf{\textcolor{blue}{\footnotesize{EXTENSIVE}}}   & \textbf{\textcolor{red}{\footnotesize{NONEXTENSIVE}}}  &\textbf{\textcolor{red}{\footnotesize{NONEXTENSIVE}}}              \\ \hline
\Espaco$\displaystyle{e.g., N^{\rho} \atop \de{\rho > 0} }$ &\textbf{\textcolor{red}{\footnotesize{NONEXTENSIVE}}} & \textbf{\textcolor{blue}{\footnotesize{EXTENSIVE}}}    &\textbf{\textcolor{red}{\footnotesize{NONEXTENSIVE}}}              \\
\rule[-2mm]{0mm}{0mm}                                      &                                                      & $\de{q = 1- 1/\rho}$                                   &                                                                   \\ \hline
\Espaco$\displaystyle{e.g., \nu^{ N^\gamma } \atop (\nu>1; }$&\textbf{\textcolor{red}{\footnotesize{NONEXTENSIVE}}} & \textbf{\textcolor{red}{\footnotesize{NONEXTENSIVE}}}  & \textbf{\textcolor{blue}{\footnotesize{EXTENSIVE}}}               \\
\rule[-2mm]{0mm}{0mm}$0<\gamma<1)$                         &                                                      &                                                        & $\de{\delta = 1/\gamma}$                                          \\ \hline
\end{tabular}
\caption{Additive and nonadditive entropic functionals and illustrative classes of systems for which the entropy is extensive. $W(N)$ is the number of admissible equally probable microscopic configurations of a system with $N$ elements; only configurations with nonvanishing occurrence probability are considered admissible.}
\label{tableBG}
\end{table}

\begin{figure}[h!]
\begin{center}
\includegraphics[width=9.2cm]{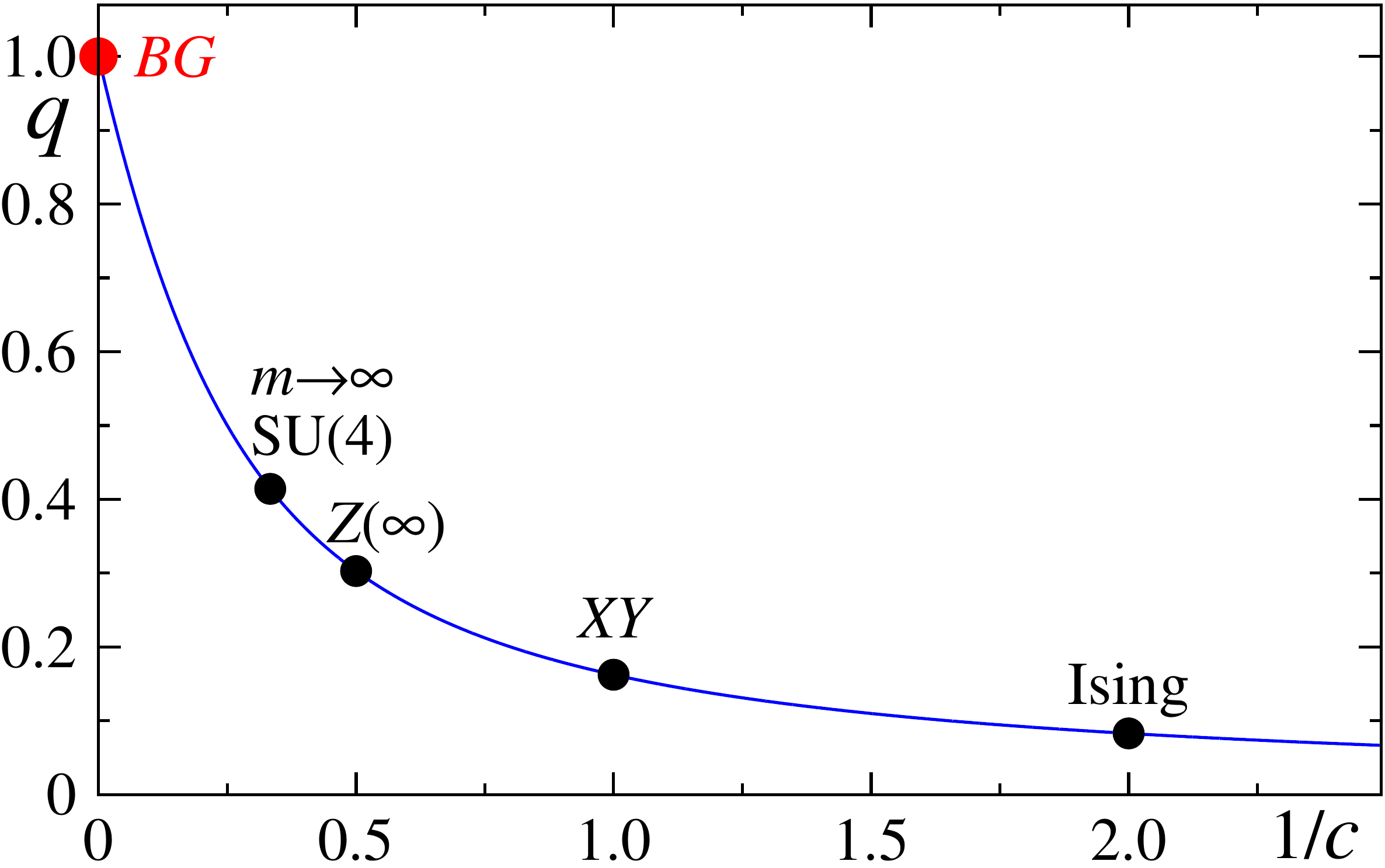}
\end{center}
\caption{The index $q$ has been determined \cite{CarusoTsallis2008} from first principles, namely from the universality class of the Hamiltonian. The values $c=1/2$ and $c=1$ respectively correspond to the Ising and XY ferromagnetic chains in the presence of transverse field at $T=0$ criticality. For other models see \cite{Alcaraz1987,Alcaraz1990}. In the $c \to\infty$ limit we recover the Boltzmann-Gibbs (BG) value, i.e.,  $q=1$. 
For arbitrary value of $c$, the subsystem {\it nonadditive} entropy $S_q$ is thermodynamically {\it extensive} for, and only for,  $q=\frac{\sqrt{9+c^2}-3}{c}$ (hence $c=\frac{6q}{1-q^2}$; 
some special values: for $c=4$ we have $q=1/2$, and for $c=6$ we have  $q=\frac{2}{\sqrt{5}+1}=\frac{1}{\Phi}$ where $\Phi$ is the {\it golden mean}).
Let us emphasize that this anomalous value of $q$ occurs {\it only} at precisely the  zero-temperature second-order quantum critical point; anywhere else the usual short-range-interaction BG behavior (i.e. $q=1$) is valid. From \cite{TsallisHaubold2015}.
}   
\label{Caruso}
\end{figure}

All the examples in Table \ref{tableBG} assume equal probabilities of the $W$ nonvanishing-probability events. What about a more generic situation?  In general, such nontrivial first-principle calculations are mathematically intractable. There are however a few exceptions. One of the most neat examples concerns the quantum critical point of a $(1+1)$-dimensional class of Hamiltonians whose continuum limit is characterized by a conformal field with {\it central charge} $c$ ($c=1/2$ corresponds to the Ising ferromagnet as well as to the axial anisotropic Heisenberg ferromagnetic model \cite{CarideTsallisZanette1983}; $c=1$ corresponds to the isotropic XY ferromagnet). It turns out that the extensive entropy of subsystems within an infinite quantum chain is $S_q$ with \cite{CarusoTsallis2008} (see also \cite{SaguiaSarandy2010}).
\begin{equation}
q= \frac{\sqrt{9+c^2}-3}{c}  \,.
\label{cequation}
\end{equation}
See Fig. \ref{Caruso}. Before going on, let us comment a point deserving further clarification The above simple illustrations in Table \ref{tableBG} focusing on how the functional form of $W(N)$ can determine an entropic functional $S(\{p_i\})$ that satisfies thermodynamic extensivity, are based, as already mentioned, on the assumption of equal probabilities. In other words, we have used the maximal value that the specific entropic functional can assume. In many cases this procedure is the correct one. It can however be wrong under some specific circumstances. To clarify this issue let us analyze with some detail the case analytically discussed in \cite{CarusoTsallis2008}. The correct formula for the index $q$ which satisfies entropic extensivity is given by Eq. (\ref{cequation}).
However, it has been analytically proved (see \cite{CalabreseCardy2004}) that, for this one-dimensional strongly-entangled quantum subsystem of size $L$, we have
\begin{equation}
\frac{S_{BG}(L)}{k} = \frac{c}{3} \ln L + \ln b + ...   \,,
\end{equation} 
where $b$ is a constant.
This expression, together with the hypothesis $S_{ BG}/k = \ln W$, implies $W(L) \sim bL^{\rho}$ with $\rho=c/3$, which would determine, {\it if} we had equal probabilities ({\it which we have not!}), $q=1-\frac{3}{c}$. The correct result is nevertheless that given in Eq. (\ref{cequation}). The erroneous value $q=1-\frac{3}{c}$ arrived by wrongly assuming equal probabilities. \\

In Section 3 we present three basic reasons why we always demand the thermodynamic entropy to be extensive. The first reason concerns what appears to be the most general form of classical thermodynamics compatible with the Legendre transform structure of usual thermodynamics. We do not assume any specific functional form for the entropy. It naturally comes out (from the Legendre structure), however, that it is extensive, i.e., that $S(N) \propto N$ or equivalently that $S(L) \propto L^d$. Along these lines, we extend usual thermodynamics (whose validity is restricted to {\it short}-range-interacting many-body systems) in order to also cover {\it long}-range-interacting many-body systems, as well as nonstandard cases such as (3+1)-dimensional and (2+1)-dimensional black holes, and the so called {\it area law} in quantum strongly entangled systems. 
The second reason concerns another strong viewpoint -- the large deviation theory --. A nontrivial example that has been numerically discussed in detail suggests, once again, that the physically admissible entropy is, for all systems, {\it extensive} in the thermodynamic sense. As a possible third reason, we end this Section by recalling and illustrating the strong similarity between the time-dependence of the entropy when the system is approaching its stationary state, and the size-dependence of the entropy when this stationary state has been attained.

In Section 4 we briefly review available results for inter-occurrence times in finance, earthquakes and genomes, as well as some other applications.
We finally conclude in Section 5. \\

\section{WHY SHOULD THE THERMODYNAMICAL ENTROPY ALWAYS BE EXTENSIVE?}
In what follows we focus on arguments yielding, as final outcome, that the thermodynamical entropy of any system must be extensive. These arguments follow along three different lines, namely thermodynamical mathematical structure, large deviation theory, and time evolution of the entropy of nonlinear dynamical systems towards their stationary states.

\subsection{Generalizing thermodynamics}

Thermodynamics is based on some very general empirical facts (which have historically led to the zeroth, first and second principles, among others). Its mathematical structure is based on Legendre transformations. Microscopically speaking, it relies on a maximum entropy principle, i.e., extremization of an entropic functional with appropriate constraints on the set of probabilities (see, for instance, \cite{Mendes1997,PlastinoPlastino1997}).

To discuss thermodynamics on general grounds we follow \cite{Tsallis2009,TsallisCirto2013} and references therein. Let us remind a typical form of the thermodynamical energy $G$ (Gibbs energy) of a generic $d$-dimensional system~\cite{Callen1985}:
\begin{eqnarray}
G(V,T,p,\mu,H, \dots) =  U(V,T,p,\mu,H,\dots)  
 - TS(V,T,p,\mu,H,\dots) \\ 
 +pV 
 -\mu N(V,T,p,\mu,H,\dots) -HM(V,T,p,\mu,H,\dots)- \cdots \,,
\label{thermodynamics}
\end{eqnarray}
where $T, p, \mu,H$ are the temperature, pressure, chemical potential, external magnetic field respectively, and $U,S,V,N,M$ are the internal energy, entropy, volume, number of particles (in turn proportional to the number of degrees of freedom), magnetization respectively. From the Legendre structure we identify three classes of variables, namely (i) those that are expected to {\it always be extensive} like $N$ itself ($S,V, N,M,\ldots$), i.e., scaling with the ($d$-dimensional) volume $V=L^d \propto N$, where $L$ is a characteristic linear dimension of the system (clearly, $V \propto A^{d/(d-1)}$, where $A$ is the $d$-dimensional area) \footnote{Within the thermodynamical Legendre-transform structure, it is of course natural that $S,V,N,M$ belong to the same class. The variable $N$ is {\it extensive} by definition. Therefore, clearly, so must also be the entropy $S$, as well as $V,M$, and similar ones.}, (ii) those that characterize the external conditions under which the system is placed ($T,p,\mu,H,\ldots$), scaling with $L^\theta$, and (iii) those that represent energies ($G,U$), scaling with $L^\epsilon$.

It trivially follows
\begin{equation}
\epsilon = \theta + d \,.
\label{thetad}
\end{equation}
If we divide Eq. (\ref{thermodynamics}) by $L^{\theta + d}$ and consider the large $L$ limit (i.e., the thermodynamical limit), we obtain
 \begin{eqnarray}
g \Bigl(\frac{T}{L^\theta},\frac{p}{L^\theta},\frac{\mu}{L^\theta},\frac{H}{L^\theta}, \dots \Bigr) =
  u\Bigl(\frac{T}{L^\theta},\frac{p}{L^\theta},\frac{\mu}{L^\theta},\frac{H}{L^\theta}, \dots \Bigr)  
 -\frac{T}{L^\theta}\,s\Bigl(\frac{T}{L^\theta},\frac{p}{L^\theta},\frac{\mu}{L^\theta},\frac{H}{L^\theta}, \dots \Bigr)       \nonumber    \\
  +\frac{p}{L^\theta}  
 -\frac{\mu}{L^\theta} \,n\Bigl(\frac{T}{L^\theta},\frac{p}{L^\theta},\frac{\mu}{L^\theta},\frac{H}{L^\theta}, \dots \Bigr) 
 -\frac{H}{L^\theta}\,m\Bigl(\frac{T}{L^\theta},\frac{p}{L^\theta},\frac{\mu}{L^\theta},\frac{H}{L^\theta}, \dots\Bigr)- \cdots \,,
\label{thermodynamics2}
\end{eqnarray}
where  $g \equiv \lim_{L\to\infty} G/L^{\theta + d}$,  $u \equiv \lim_{L\to\infty} U/L^{\theta +d}$, $s \equiv \lim_{L\to\infty} S/L^d$, $n \equiv \lim_{L\to\infty} N/L^d$, $m \equiv \lim_{L\to\infty} M/L^d$.

The correctness of the scalings appearing in this equation has been profusely verified in the literature for both short- and long-range interacting classical thermal~\cite{JundKimTsallis1995,Grigera1996,CannasTamarit1996,SampaioAlbuquerqueMenezes1997,AnteneodoTsallis1998,CurilefTsallis1999,AndradePinho2005,BinekPolisettyHeMukherjeeRajeshRedepenning2006}, diffusive~\cite{diffusion}, and geometrical (percolative) systems~\cite{RegoLucenaSilvaTsallis1999,FulcoSilvaNobreRegoLucena2003}. The case of short-range interactions corresponds to the standard thermodynamical systems (e.g., a real gas, a simple metal), and we have $\theta =0$ (i.e., we recover the usual {\it intensive} variables such as $T$, $p$, $\mu$, $H$.), and $\epsilon = d$ (i.e., we recover the usual {\it extensive} energy variables such as $G$, $U$, etc). This is the case that is found in the textbooks of thermodynamics (see, for instance \cite{Callen1985}). The case of long-range interactions is more subtle. This case is conveniently discussed by introducing a variable such as the following one \cite{JundKimTsallis1995,Grigera1996,CannasTamarit1996,SampaioAlbuquerqueMenezes1997,AnteneodoTsallis1998,CurilefTsallis1999,AndradePinho2005,BinekPolisettyHeMukherjeeRajeshRedepenning2006,diffusion,RegoLucenaSilvaTsallis1999,FulcoSilvaNobreRegoLucena2003}:

\begin{equation}
{\tilde N} \equiv \frac{N^{1-\alpha/d}-\alpha/d}{1-\alpha/d} \;\;\;(\alpha \ge 0)\,,
\end{equation}
where $\alpha$ characterizes the range of a two-body interaction within a classical $d$-dimensional system (assuming that the potential is integrable at the origin, and asymptotically decays as $1/r^\alpha$ at long distances $r$).

\begin{figure}[h!]
\centering
 \resizebox{0.90\columnwidth}{!}{\includegraphics{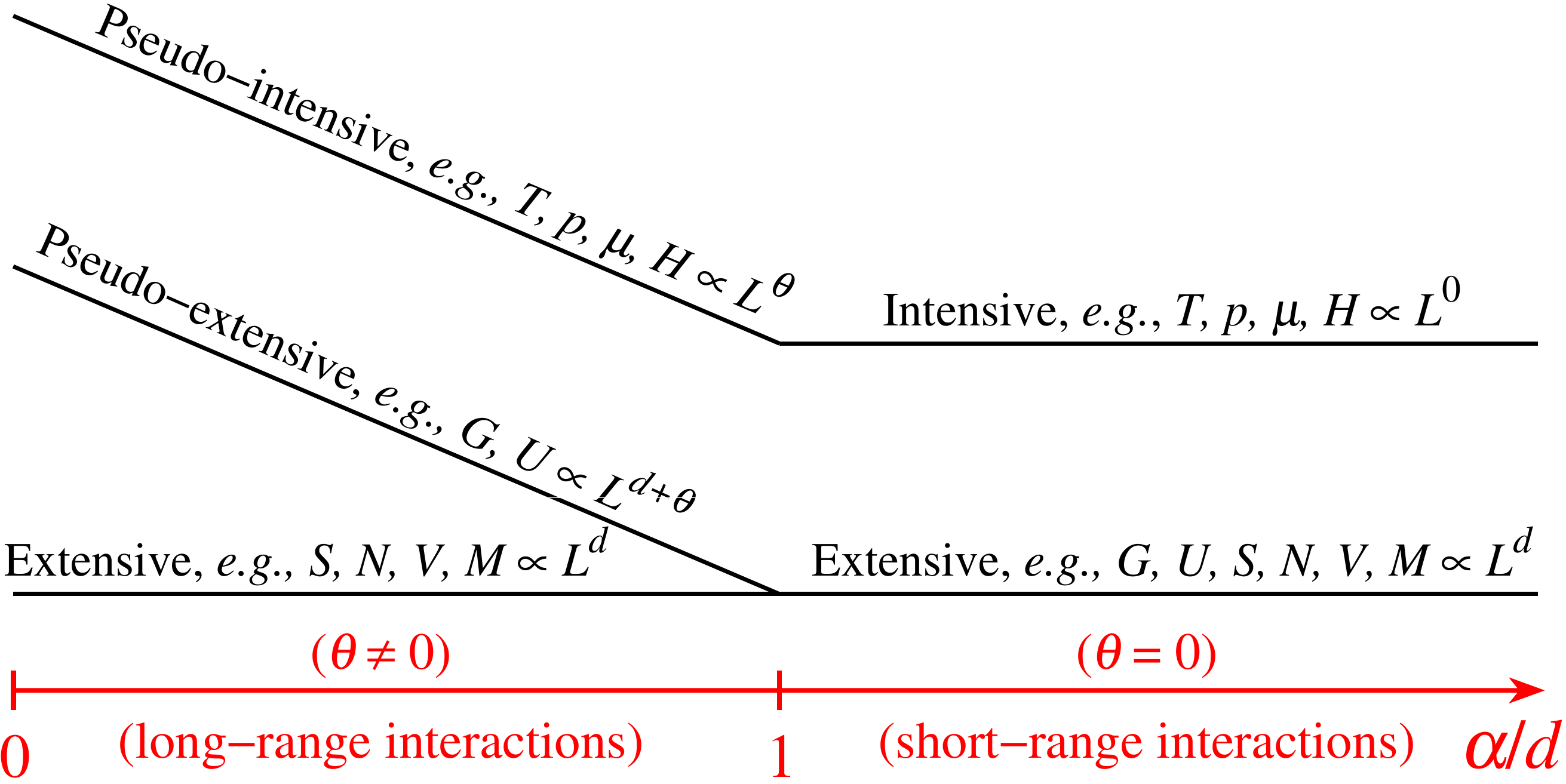} }
 \vspace{0.5cm}
\caption{
Representation of the different scaling regimes of the Eq.~\eqref{thermodynamics2} for classical $d$-dimensional systems.
For attractive long-range interactions (\emph{i.e.}, $0 \leq \alpha/d \leq 1$, $\alpha$ characterizes the interaction range in a potential with the form $1/r^{\alpha}$)
we may distinguish three classes of thermodynamic variables, namely,
those scaling with $L^{\theta}$, named \emph{pseudo-intensive} ($L$ is a characteristic linear length, $\theta$ is a system-dependent parameter),
those scaling with $L^{d+\theta}$, the \emph{pseudo-extensive} ones (the energies),
and those scaling with $L^{d}$ (which are always extensive).
For short-range interactions (\emph{i.e.}, $\alpha > d$)
we have $\theta=0$ and the energies recover their standard~$L^{d}$ extensive scaling,
falling in the same class of~$S$, $N$, $V$, etc, 
whereas the previous pseudo-intensive variables become truly intensive ones (independent of $L$);
this is the region, with two classes of variables, that is covered by the traditional textbooks of thermodynamics. From \cite{Tsallis2009}.
}
\label{Fig:Pseudo}
\end{figure}

As we see, if $\alpha /d >1$ (short-range interactions), we have, in the $N \to \infty$ limit, a constant ${\tilde N}$, thus recovering once again standard thermodynamics. But for $0 \le \alpha /d \le 1$ (short-range interactions), we have, in the $N \to \infty$ limit, that ${\tilde N}$ scales in a nontrivial manner, namely ${\tilde N} = \ln N$ if $\alpha/d=1$, and ${\tilde N} \sim N^{1-\alpha/d}/(1-\alpha/d)$ if $0 \le \alpha/d<1$ (for the particular case $\alpha=0$ we recover the usual mean field scaling ${\tilde N} = N$). For $0 \le \alpha/d \le 1$, we have that
\begin{equation}
\theta=d-\alpha \,,
\end{equation}
and
\begin{equation}
\epsilon=2d-\alpha \,,
\end{equation}
which means that $(T,p,\mu,H)$ are non-intensive, and $(G,U)$ are superextensive\footnote{Quantum $d$-dimensional systems are slightly different. For them we expect the usual thermodynamical scalings to hold for $\alpha > \alpha_c(d)$, where typically $\alpha_c(d) >d$ (see, for example, \cite{BorlandMenchero1999,BorlandMencheroTsallis2000}), whereas anomalous scalings emerge for $0 \le \alpha \le \alpha_c(d)$.}. We verify that, remarkably enough, $(N,S,V,M)$ remain extensive in {\it all} cases, i.e., $\forall \,\alpha/d$. See Fig. \ref{Fig:Pseudo}. 
These peculiar scalings are a consequence from the fact that such potentials are not integrable, i.e., from the fact that the integral $\int_{\textrm{constant}}^\infty dr \,r^{d-1}\,r^{-\alpha}$ diverges, and therefore the BG canonical partition function itself diverges. In his 1902 book {\it Elementary Principles in Statistical Mechanics} \cite{Gibbs1902},  Gibbs himself emphatically points out that whenever the partition function diverges, the BG theory can not be used (in his words ``the law of distribution becomes illusory"). As an illustration of his remark he refers specifically to the case of Newtonian gravitation (i.e., $d=3$ and $\alpha=1$)
\footnote{From the microscopic (classical) dynamical point of view, this anomaly is directly related to the fact that the entire Lyapunov spectrum vanishes in the $N \to \infty$ limit, which can impeach ergodicity (see \cite{AnteneodoTsallis1998,CampaGiansantiMoroniTsallis2001} and references therein).
This type of difficulty is also present, sometimes in an even more subtle manner, in various quantum systems (the single hydrogen atom constitutes, among many others, an elementary such example; indeed its BG partition function diverges due to the accumulation of electronic energy levels just below the ionization energy).}.

In addition to the above long-range-interacting classical systems, other ones exist which also have intriguing aspects in what concerns their thermodynamics. Such is the case of black holes, whose entropy is being currently discussed since several decades.
Indeed, since the pioneering works of Bekenstein~\cite{Bekenstein1973a,Bekenstein1973b} and Hawking~\cite{Hawking1974a,Hawking1974b},  it has become frequent in the literature the (explicit or tacit) acceptance that the black-hole entropy is ano\-ma\-lous in the sense that it can violate thermodynamical extensivity~\cite{tHooft1985,tHooft1990,Susskind1993,Maddox1993,Srednicki1993,StromingerVafa1996,MaldacenaStrominger1998,DasandShankaranarayanan2006, BrusteinEinhornYarom2006,BorstenDahanayakeDuffEbrahimRubens2009,Padmanabhan2009,Casini2009,BorstenDahanayakeDuffMarraniRubens2010,Corda2011,KolekarPadmanabhan2011,Saida2011}.
We frequently read and hear claims that the entropy of the black hole is proportional to the area of its boundary instead of being proportional to the black-hole volume.

For a Schwarzschild $(3+1)$-dimensional black hole, the energy scales like the mass $M_{bh}$ (where bh stands for {\it black hole}), which in turn scales with $L$ \cite{generalrelativity, generalrelativity_2, generalrelativity_3}, hence $\epsilon =1$, hence, using Eq. (\ref{thetad}),
\begin{equation}
\theta = 1-d\,.
\label{1d}
\end{equation}
{\it If} the black hole is physically identified with its event horizon surface, then it is to be considered as a genuine $d=2$ system, then $\theta =-1$, which precisely recovers the usual Bekenstein-Hawking (BH) scaling $T \propto 1/L \propto 1/M_{bh}$.
\emph{If} however the black hole is to be considered as a genuine $d=3$ system (which can make sense given that the corresponding space-time is (3+1)-dimensional), then $\theta =-2$, i.e., $T$ scales like $1/L^2 \propto 1/M_{bh}^2$, in variance with the BH scaling. This is a manner for understanding why such a puzzle exists since decades related to the entropy of a black hole. Let us be somewhat more specific. Wide and physically meaningful evidence (e.g., the holographic principle) exists that the BG entropy (for quantum systems, also referred to as {\it von Neumann} entropy) $S_{BG} \equiv k_B\ln W \propto L^2$, and more generally that $S_{BG} \equiv -k_B Tr \rho \ln \rho \propto L^2$, $W$ being the total number of internal configurations,
and $\rho$ being the density matrix.
For strongly quantum-entangled $d$-dimensional systems we similarly have what is currently referred to as the {\it area law} \cite{EisertCramerPlenio2010}, i.e., the fact that $S_{BG} \equiv -k_BTr \rho \ln \rho$ frequently scales with $L^{d-1}$ for $d>1$, and with $\ln L$ for $d=1$, instead of scaling, for $d \ge 1$, with $L^d$. This fact also generates a closely related intriguing question. The above remarks might be considered the heart of the ongoing discussion for the entropy of a black hole. Indeed, {\it if} the system is to be physically considered a $(d-1)$-dimensional one, then the (additive) entropy $S_{BG}$ certainly is to be identified as its thermodynamical entropy. But {\it if} the system is to be physically considered a $d$-dimensional one, then $S_{BG}$ can {\it not} be identified as its thermodynamical entropy, and, as we can see, a nonadditive entropic functional is needed  to play that role \cite{TsallisCirto2013}. It is shown in \cite{TsallisCirto2013} that, under the assumption of equal probabilities, the nonadditive entropic functional to be used is the so called $\delta$-entropy, $S_\delta$, with a special value of the $\delta$ index, namely $\delta=d/(d-1)$ ($d>1$). If, however, the equal-probabilities hypothesis is not verified for these anomalous systems, then the nonadditive functional to be used could be (up to our present knowledge) $S_\delta$ with some other value of $\delta$, or it could be something else, for example once again the functional $S_q$ itself with a specific value of the index $q$.

Let us emphasize the above. {\it If} we are to consider the Schwarzschild $(3+1)$ black hole 
as a genuine~$d=2$ system, then $S_{BG} = k_B \ln W \propto L^2$ corresponds indeed to the (extensive) thermodynamical entropy $S$, the BH scaling $T \propto 1/M_{bh}$ is to be expected, and there are no controversial or intriguing facts to be further analyzed.
{\it If} however, this black hole is to be considered a genuine~$d=3$ system, then~$S_{BG}$ {\it can not be the thermodynamical entropy~$S$}, since the latter must scale like~$L^3$ whereas $S_{BG}$ scales like $L^2$. Within this standpoint, a crucial question then arises, namely,  {\it what is then the microscopic mathematical expression of the thermodynamical entropy $S$ of this 3-dimensional system?}
We provide in \cite{TsallisCirto2013} a thermodynamically admissible answer to this important question.

The $(2+1)$-dimensional ``black hole" has been discussed as well~\cite{carlip, carlip_2, carlip_3}.
It has been shown that the energy scales like~$L^2$, hence $\epsilon = 2$ and, 
using Eq. (\ref{thetad}) once again, 
\begin{equation}
\theta=2-d \,.
\end{equation}
This case provides an event horizon which is one-dimensional.
{\it If}, due to this fact, this black hole is to be considered a genuine $d=1$ system,
then $\theta = 1$,  which corresponds to the $(2+1)$ version of BH scaling, i.e., $T\propto L\propto M_{bh}^{1/2}$.
Indeed, this is precisely the scaling that has been obtained~\cite{carlip,carlip_2,carlip_3} for this simplified system.
{\it If}, however, this black hole is to be considered as a $d=2$ system, we have $\theta=0$, and, in this case, $T$ is expected to be an intensive variable.
Consistently, {\it if} we assume the system to be a $d=1$ one, then clearly $S_{BG}$ plays the role of its thermodynamical entropy, since~$S_{BG}\propto L$ (as obtained in~\cite{carlip,carlip_2,carlip_3}).
But, similarly to the $(3+1)$ case discussed above, {\it if} we consider it to be a $d=2$ one, then once again a nonadditive entropic functional is needed to play the thermodynamical role.

From a historical perspective, we observe that, strangely enough, Gibbs's crucial warning about the partition function being divergent in some anomalous cases, and the dramatic theoretical features to which this is related, are overlooked in most textbooks.
Similarly, the thermodynamical violation related to the area law frequently is, somehow, not taken that seriously. Indeed, not few authors seem inclined to consider that, for such complex systems, the entropy is not expected to satisfy thermodynamical extensivity.
In contrast, however, various physical and mathematical facts exist which reveal such standpoint as kind of bizarre.
One of the main goals of \cite{TsallisCirto2013} is to address this important issue and develop a path along which the difficulty can be overcome.
The fact (repeatedly illustrated in various manners for strongly entangled systems, black holes and, generically speaking, for systems satisfying the above mentioned area law) that the Boltzmann\--Gibbs\--von Neumann ({\it additive}) entropy is {\it not} proportional to the volume $L^d$ precisely shows that, for such strongly correlated systems
(hence the total number of admissible states in phase space is sensibly reduced), {\it the thermodynamical entropy can not be identified with the usual (additive) BG entropic functional but with a substantially different (nonadditive) one}.
An argument reinforcing the correctness of using nonadditive entropic forms in order to re-establish the entropic extensivity of the system can be found in the results achieved by Hanel and Thurner~\cite{HanelThurner2011a, HanelThurner2011b,HanelThurnerGellMann2014} by focusing on the Khinchine axioms and on complex systems with surface-dominant statistics. See also \cite{Tempesta2015}. 

\subsection{Towards a generalized large-deviation theory}

A recent result exists~\cite{RuizTsallis2012,Touchette2013,RuizTsallis2013}, related to the so called Large Deviation Theory (LDT) in theory of probabilities, which also is consistent with the extensivity of the entropy, even in the presence of strong correlations between the elements of the system.

The {\it $q$-exponential function} $e_q^z \equiv [1 + (1 - q) z]^{\frac{1}{1-q}}$ ($e_1^z = e^z$) (and its associated $q$-Gaussian \cite{reminder}) has already emerged in a considerable amount of nonextensive and similar systems (see \cite{GellMannTsallis2004,AnteneodoTsallis1998,TsallisAnjosBorges2003,TirnakliJensenTsallis2011,ZandTirnakliJensen2015,PluchinoRapisardaTsallis2007,TamaritCannasTsallis1998,AnteneodoTsallis1997,TirnakliTsallisLyra1999,RodriguezSchwammleTsallis2008,AndradeSilvaMoreiraNobreCurado2012,PlastinoPlastino1995,pedron,AnteneodoTsallis2003,UpadhyayaRieuGlazierSawada2001,ThurnerWickHanelSedivyHuber2003,DanielsBeckBodenschatz2004,BurlagaVinas2005,DouglasBergaminiRenzoni2006,LutzRenzoni2013,CombeRichefeuStasiakAtman2015,LiuGoree2008,CMS,TirnakliBeckTsallis2007,TirnakliTsallisBeck2009,TsallisTirnakli2010,NobreRegoMonteiroTsallis2011,NobreRegoMonteiroTsallis2012}  among others), as the appropriate generalization of the exponential one (and its associated Gaussian) (with regard to the nonlinear quantum equations introduced in \cite{NobreRegoMonteiroTsallis2011}, see also \cite{CostaAlmeidaFariasAndrade2011,NobreRegoMonteiro2012,Mazharimousavi2012,PlastinoTsallis2013,CostaAlencarSkagerstamAndrade2013,RegoMonteiroNobre2013a,ToranzoPlastinoDehesaPlastino2013,RegoMonteiroNobre2013b,CurilefPlastinoPlastino2013,PenniniPlastinoPlastino2014,CostaBorges2014}). Therefore, it appears as rather natural to conjecture that, in some sense that remains to be precisely defined, the LDT expression $e^{-r_1N}$ becomes generalized into something close to $e_q^{-r_q N}$ ($q \in {\cal R}$), where the generalized rate function $r_q$ should be some generalized entropic quantity {\it per particle}.

\begin{figure}[h!]
\begin{center}
\includegraphics[width=14cm]{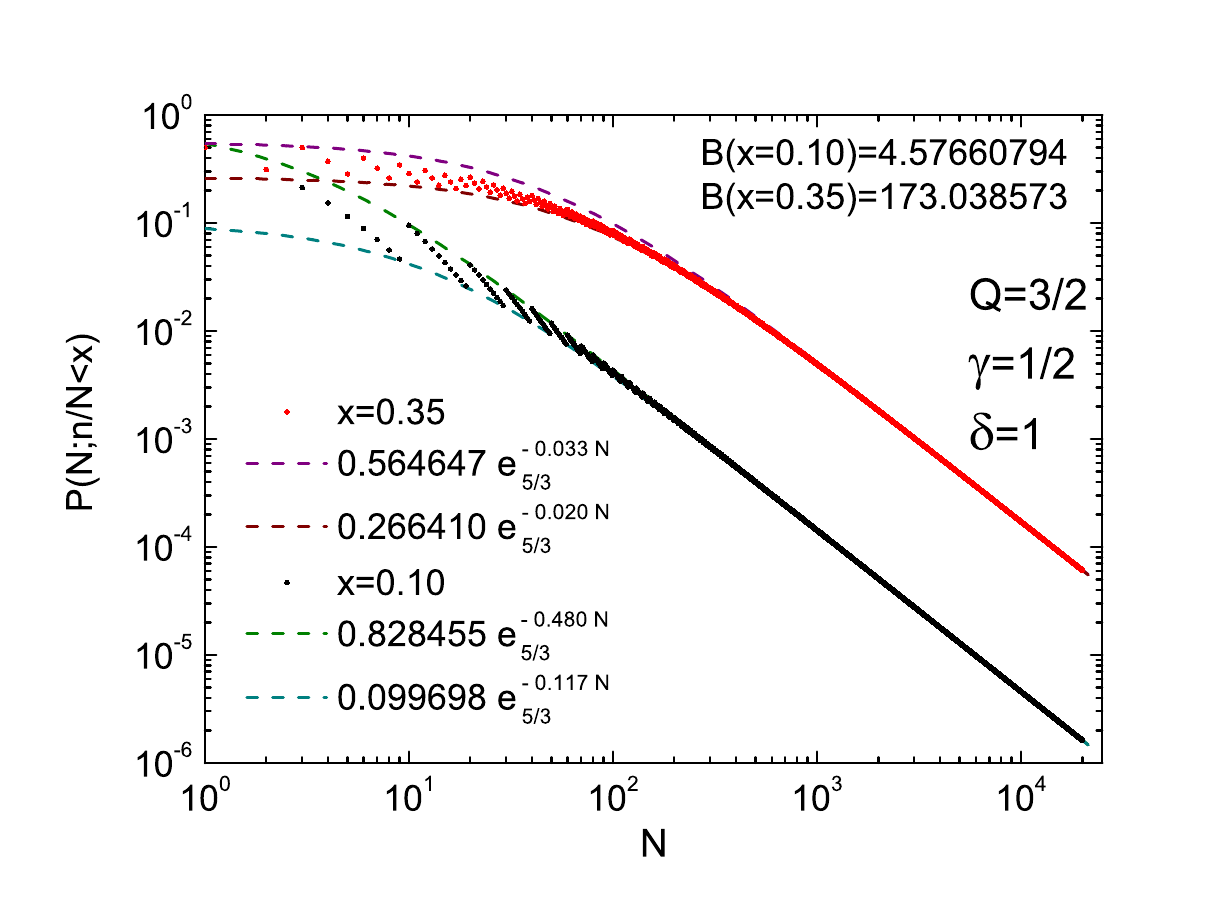}
\end{center}
\vspace{-1cm}
\caption{Comparison of the numerical data (dots) with $a(x)e_q^{-r_q N}$. Two values for $x$, namely $x=0.10$ and $x=0.35$, have been illustrated here. From \cite{RuizTsallis2013}.
}
\label{comparison}
\end{figure}

\begin{figure}[h!]
\begin{center}
\includegraphics[width=14cm]{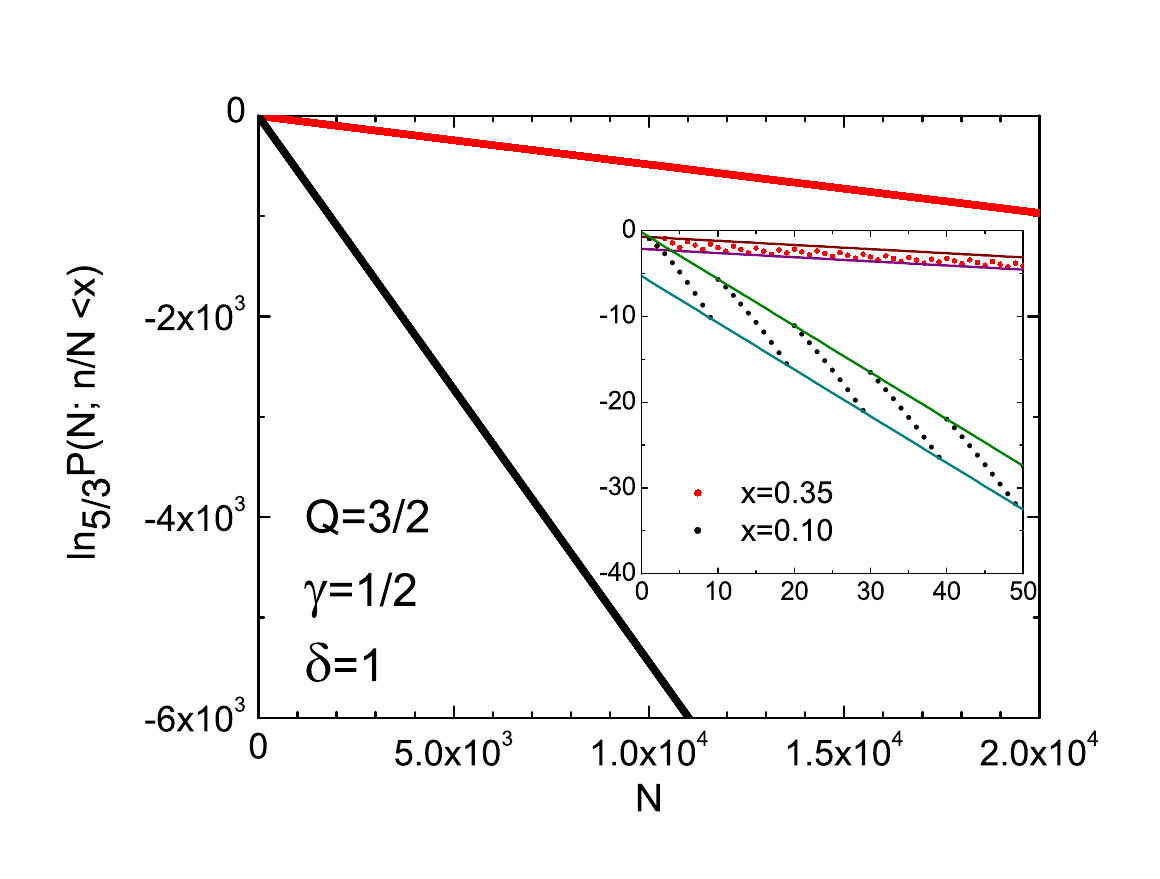}
\end{center}
\vspace{-1cm}
\caption{The same data of Fig. \ref{comparison} in ($q$-log)-linear representation. Let us stress that the {\it unique} asymptotically-power-law function which provides straight lines {\it at all scales} of a ($q$-log)-linear representation is the $q$-exponential function. The inset shows the results corresponding to $N$ up to 50. From \cite{RuizTsallis2013}.
}
\label{comparisonqlog}
\end{figure}

Let us stress a crucial point: we are {\it not} proposing for long-range interactions, and other nonstandard systems, something like $e_q^{-r_q N^\eta}$ with $\eta \ne 1$, but we are expecting instead $\eta =1$, i.e., the {\it extensivity} of the total $q$-generalized entropy to still hold. As we can see in Figs. \ref{comparison} and \ref{comparisonqlog}, this important assumption indeed is indeed verified in the model, characterized by $(Q,\gamma,\delta)$ with $Q \ge 1$, that we numerically studied in \cite{RuizTsallis2012,RuizTsallis2013}. The index $q \ge 1$ satisfies
\begin{equation}
\frac{1}{\gamma(q-1)}= \frac{2}{Q-1}-1 \,.
\end{equation}
This result constitutes a strong indication that, consistently with other results available in the literature (see, for instance, \cite{GellMannTsallis2004,Tsallis2009,JundKimTsallis1995,diffusion,RegoLucenaSilvaTsallis1999}), the total entropy might remain {\it extensive} (i.e., thermodynamically admissible) even in nonstandard cases where the BG entropy fails to be extensive. Any analytical results along these or similar lines would obviously be highly interesting and welcome.


\subsection{Time evolution of the entropy of nonlinear dynamical systems}

\begin{figure}[h!]
\begin{center}
\includegraphics[width=11cm]{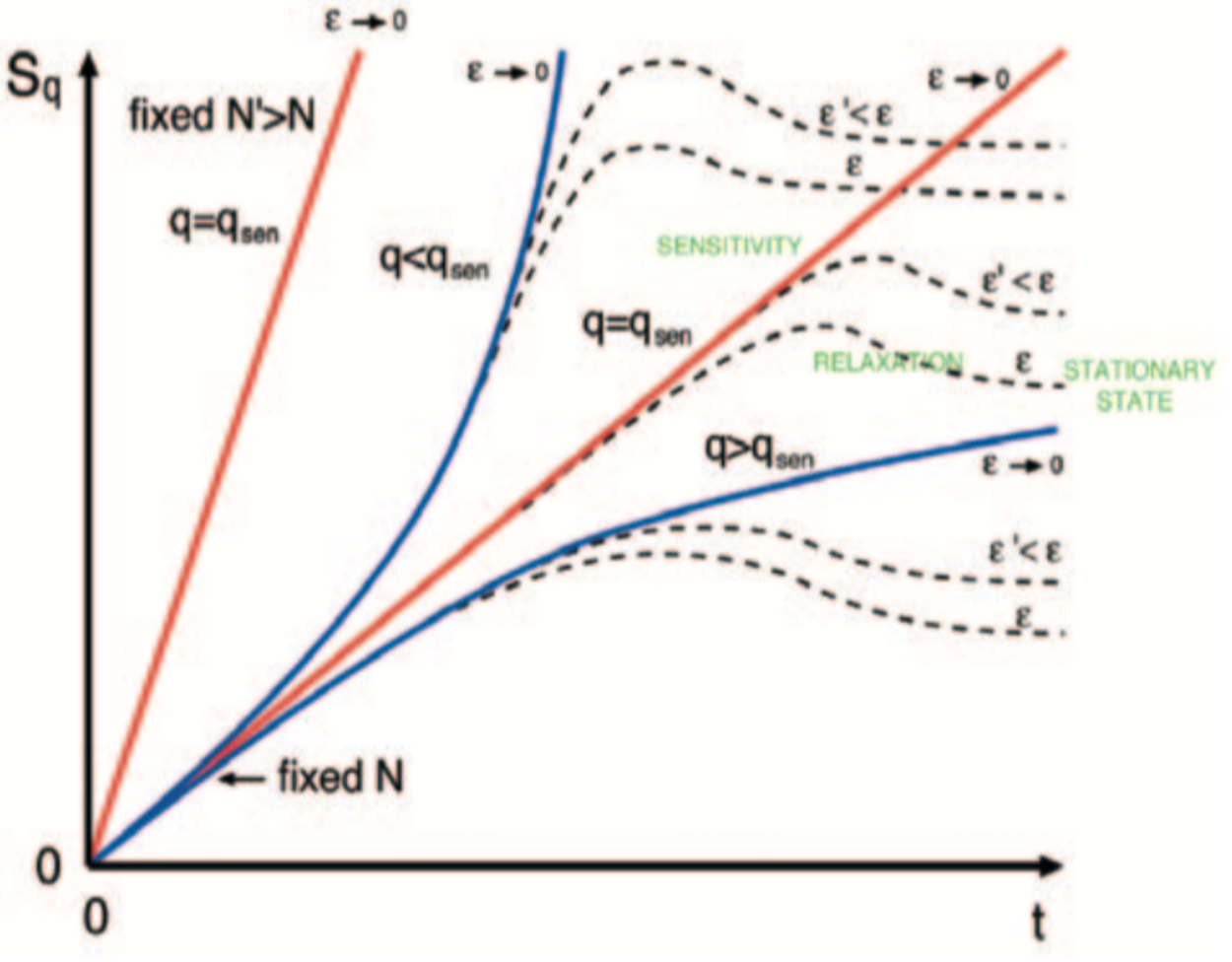}
\end{center}
\caption{The conjectural dependence of $S_q$ as a function of $(N,t)$.  A special value of $q$ (noted $q_{ent}$ in the present paper, and noted $q_{sen}$ in the original figure of \cite{TsallisGellMannSato2005b} here reproduced) is expected to generically exist such that $S_{q_{ent}}(N,t) \propto Nt$ for $N>>1$ and $t>>1$ if $\epsilon  \to 0$, where $\epsilon$ is the degree of fine-graining. The limit $\epsilon  \to 0$ corresponds to an infinitely fine binning of a (continuous) classical space-phase. This conjecture is, for $\epsilon >0$, consistent with $\lim_{t \to\infty} S_{q_{ent}}(N,t) \propto  S_{q_{ent}}(N,0<t<<t_{saturation})  \propto N$, where $t_{saturation}$ is the characteristic time for which $S_{q_{ent}}(N,t)$ roughly achieves, for increasing $t$, the value $S_{q_{ent}}(N,\infty)$. See \cite{TsallisGellMannSato2005b} for further details.
}
\label{conjecture}
\end{figure}

A further indication we can refer to is the analogy with the time $t$ dependence of the entropy of simple nonlinear dynamical systems, e.g., the logistic map $x_{t+1}=1-ax_t^2$.
Indeed, for the parameter values of $a$ for which the logistic map has {\it positive} Lyapunov exponent $\lambda_1$ (i.e., strong chaos and ergodicity; for example, for $a=2$), we verify (under appropriate mathematical limits) that $K_1\equiv \lim_{t\to\infty}\frac{S_{BG}(t)}{t} =\lambda_1$. In contrast, for parameter values where the Lyapunov exponent {\it vanishes} (namely, for weak chaos and breakdown of ergodicity; for example, at the Feigenbaum point $a=1.40115518909...$), it is a nonadditive entropy ($S_q$, discussed below) the one which grows linearly with $t$ (see~\cite{TsallisPlastinoZheng1997,BaldovinRobledo2004_2, BaldovinRobledo2004_3, BaldovinRobledo2004_4, BaldovinRobledo2004_5, BaldovinRobledo2004_6, BaldovinRobledo2004_7, BaldovinRobledo2004_8, BaldovinRobledo2004_9, BaldovinRobledo2004_10,FuentesTsallisSato2011} and references therein), and consistently provides a generalized Pesin-like identity. More precisely, $K_q\equiv \lim_{t\to\infty}\frac{S_{q}(t)}{t} =\lambda_q$, with $\lambda_q=1/(1-q)$, and $q=0.244487701341282...$ (1,018 exact digits are actually known).

If we take into account that, in many (if not all) such dynamical systems, $t$ plays a role analogous to $N$ in thermodynamical systems (see \cite{TsallisGellMannSato2005b,LuqueLacasaRobledo2012}, and Fig. \ref{conjecture}), we have here one more suggestive indication which aligns with the extensivity of the entropy for both simple and complex systems.  To be more specific, if we have a nonlinear dynamical system with ${\bar N}$ positive Lyapunov exponents $\{\lambda_1^{(k)}\}$, the Pesin equality stands basically as follows:
\begin{equation}
K_1\equiv \lim_{t\to\infty}\frac{S_{BG}(t)}{t} =\sum_{k=1}^{{\bar N}} \lambda_1^{(k)} \,.
\end{equation}
If all of these Lyapunov vanish, we hopefully have, for a wide class of weakly chaotic systems, something like the following $q$-generalization:
\begin{equation}
K_{q_{ent}}\equiv \lim_{t\to\infty}\frac{S_{q_{ent}}(t)}{t} =\sum_{k=1}^{{\bar N}}\lambda_{q_k}^{(k)} \,,
\end{equation}
where $q_{ent}$ could be determined through (see Eq. (26) of \cite{AnanosBaldovinTsallis2005})
\begin{equation}
\frac{1}{1-q_{ent}}=\sum_{k=1}^{{\bar N}} \frac{1}{1-q_k} \,,
\label{sum}
\end{equation}
where the set $\{ q_k\}$ is defined through the sensitivities $\{\xi_k\}$ to initial conditions along the various directions of nonlinear dynamical expansions, $\xi_k \simeq e_{q_k}^{\lambda_{q_k}^{(k)} \,t}  \propto t^{\frac{1}{1-q_k}} \;\;(t \to\infty)$. If at least one of the $\{ \lambda_{1}^{(k)} \}$ is positive, we have that at least one of the $\{ q_k \}$ equals one, hence $q_{ent}=1$. Another interesting particular instance appears if all ${\bar N}$ directions are degenerate (in the sense that the corresponding random variables are exchangeable, hence $q_k=q_1 \,, \forall k$), then $\frac{1}{1-q_{ent}}=\frac{{\bar N}}{1-q_1}$; this relations implies $q_{ent}=1$ in the limit ${\bar N} \to\infty$, even if $q_1<1$. Finally, another most interesting particular case is also possible, namely that the sum in (\ref{sum}) converges onto a finite positive value; we then have $q_{ent} <1$. So, if we consider a time-evolving system constituted by $N$ strongly correlated particles, it seems  plausible that, in many cases, we would have, for large $t$ and large $N$, a ratio $\frac{S_q(N,t)}{Nt}$ that would  be finite for one and the same special value of $q$, namely $q_{ent}<1$. 

\section{INTER-OCCURRENCE TIMES IN FINANCE, EARTHQUAKES AND GENOMES}

Let us now focus on an interesting universal property, shared by many natural, artificial and social systems constituted by strongly-correlated elements, namely the distribution of specific inter-occurrence times. We will specially review some available results in finance \cite{LudescherTsallisBunde2011,LudescherBunde2014}, earthquakes \cite{AntonopoulosMichasVallianatosBountis2014}, and biology \cite{BogachevKayumovBunde2014}, among others \cite{TirnakliJensenTsallis2011,ZandTirnakliJensen2015,CelikogluTirnakliQueiros2010}. We will not include in the present occasion the discussion of long-range-interacting Hamiltonian classical systems \cite{PluchinoRapisardaTsallis2007,NobreTsallis1995,LucenaSilvaTsallis1995,PluchinoRapisarda2007,PluchinoRapisardaTsallis2008,FigueiredoRochaAmato2008,PluchinoRapisardaTsallis2009,FigueiredoRochaAmato2009,CirtoAssisTsallis2014,ChristodoulidiTsallisBountis2014} which surely deserve a detailed study by themselves (in the present context we do not refer to systems whose elements short-range-interact, but only to those whose elements long-range-interact). 

\subsection{Inter-occurrence times in financial time series}

Let us consider a time series of financial prices $\{ P_t \}$. {\it Returns} are defined as follows:
\begin{equation}
r_t \equiv \ln \frac{P_t}{P_{t-1}} \simeq \frac{P_t - P_{t-1}}{P_t} \,.
\end{equation}
Return variables are convenient, hence quite popular. Indeed, they neutralize possible inflation in the prices; also, they reflect very simply up-down oscillations of {\it relative} prices. It is well known that the distributions of returns typically are well fitted by $q$-Gaussians, i.e., $P(r) \propto e_{q_r}^{-\beta_r r^2}$, with a value of $q_r>1$ which monotonically decreases towards unity for increasingly large time intervals between successive prices. Indeed, for very large time steps, all correlation is lost and the return distributions become well fitted by Gaussians. 
In the present review we follow the content of \cite{LudescherTsallisBunde2011}.

\begin{figure}[h!]
\begin{center}
\includegraphics[width=8cm]{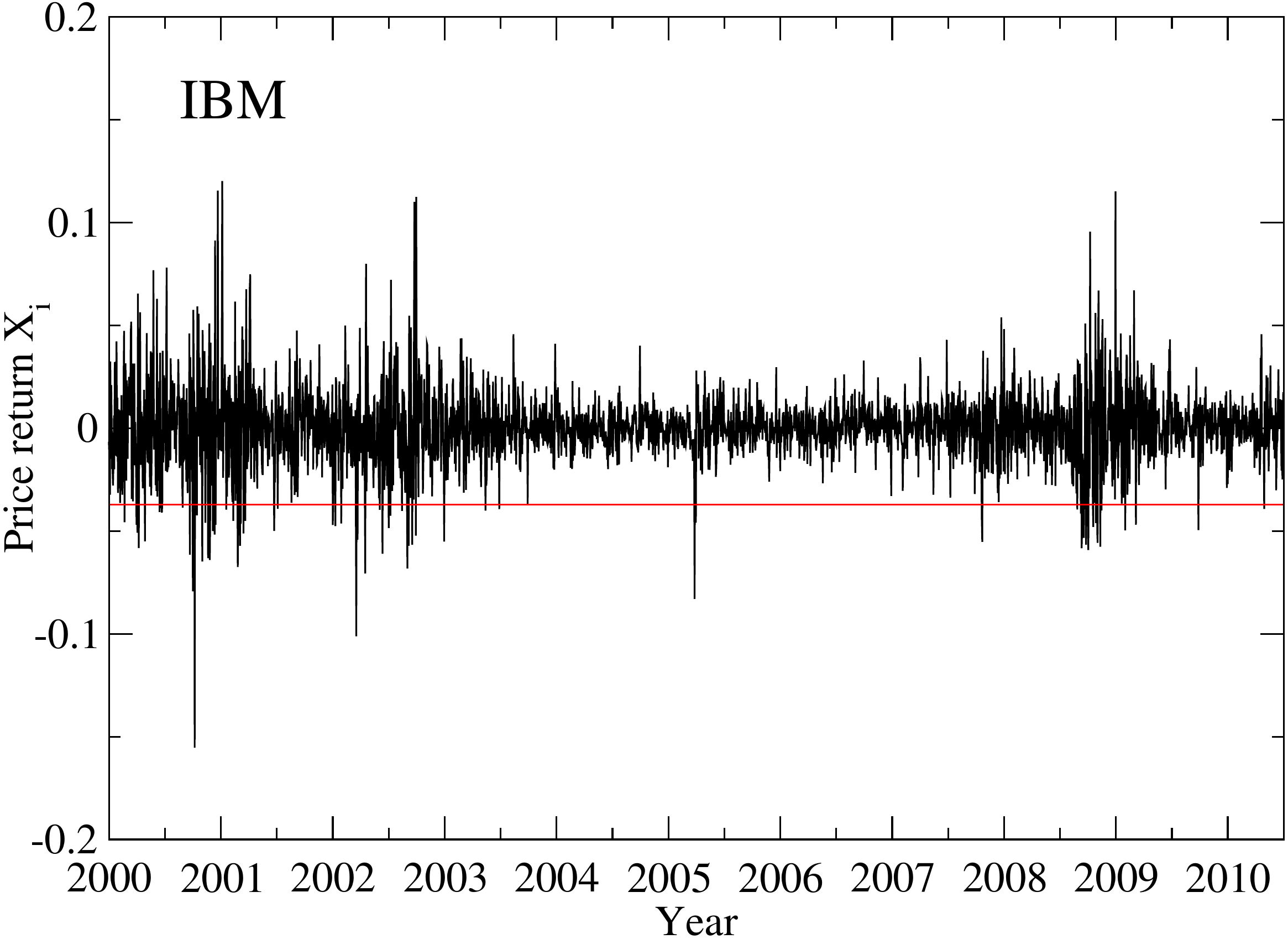}
\includegraphics[width=11.5cm]{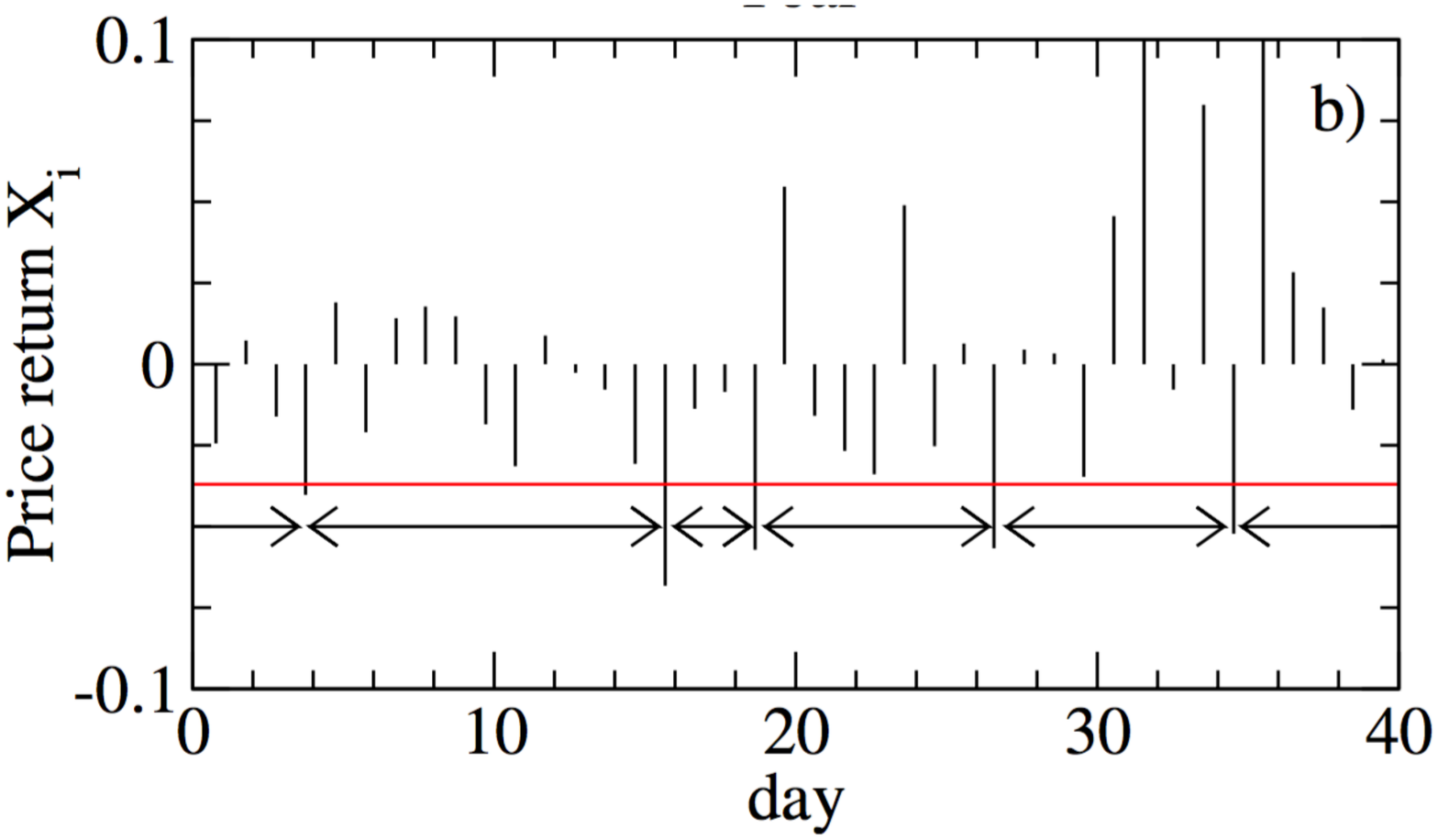}
\end{center}
\vspace{-2.0cm}
\caption{Illustration of the relative daily price returns $X_i$ of the IBM stock between (a) 2000 and 2010, and (b) August 29 and October 23, 2002. 
The red line shows the threshold $Q\simeq-0.037$, which corresponds to an average interoccurence time of $R_Q=70$. In (b) the inter-occurrence times are indicated by arrows. From \cite{LudescherTsallisBunde2011}.
}
\label{IBM-Series}
\end{figure}

\begin{figure}[h!]
\begin{center}
\vspace{-3.0cm}
\includegraphics[width=16cm]{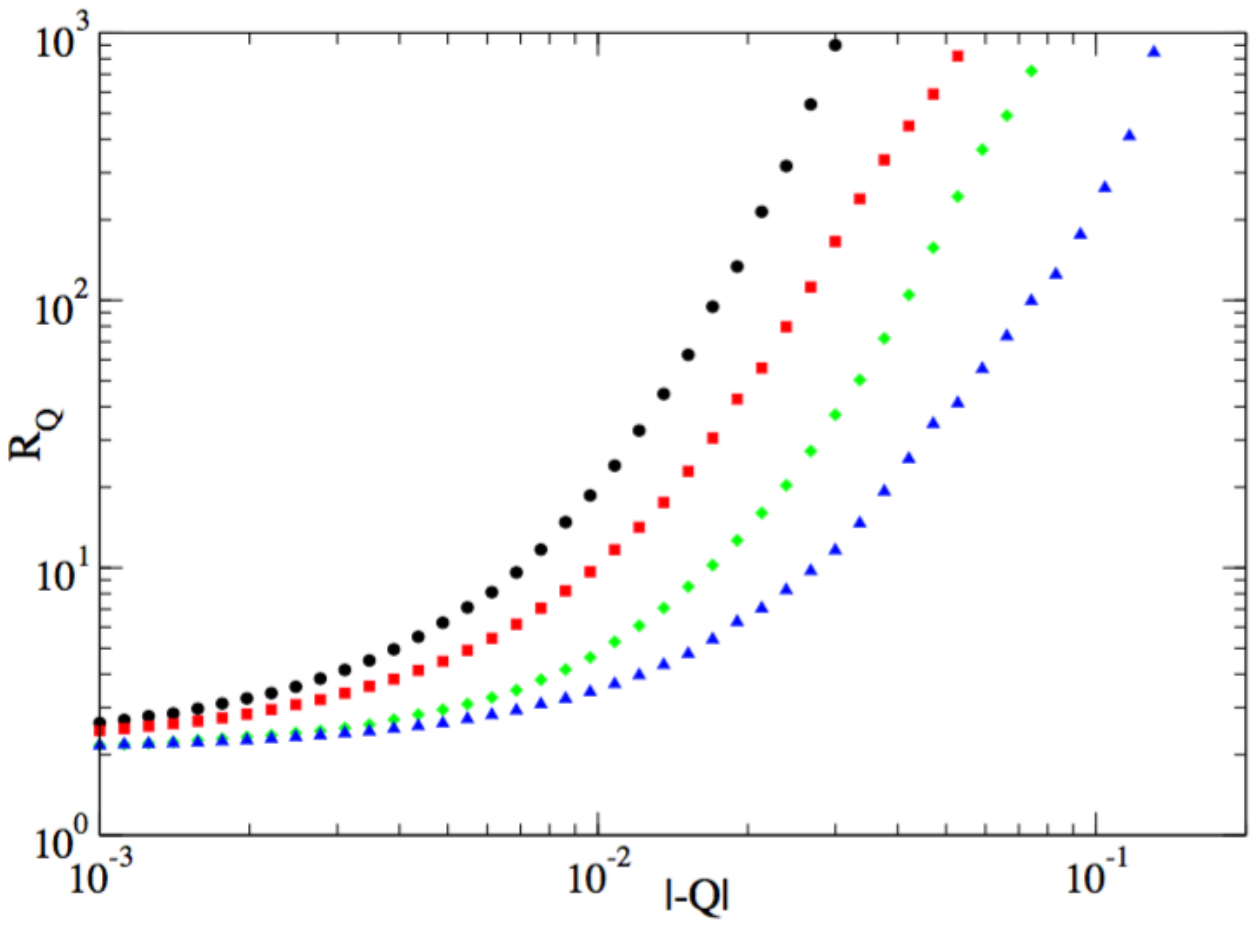}
\end{center}
\vspace{-8.0cm}
\caption{The mean inter-occurrence time $R_Q$ versus the absolute value of the loss threshold $-Q$, for the exchange rate US Dollar against British Pound, the index S$\&$P500, the IBM stock, and crude oil (WTI), from left to right. It can be verified that roughly $R_Q \simeq e_{q_R}^{\kappa\,|-Q|}$} with $q_R<1$ and $\kappa >0$. From \cite{LudescherTsallisBunde2011}.
\label{RQ}
\end{figure}

\begin{figure}[h!]
\vspace{-6.5cm}
\begin{center}
\includegraphics[width=16cm]{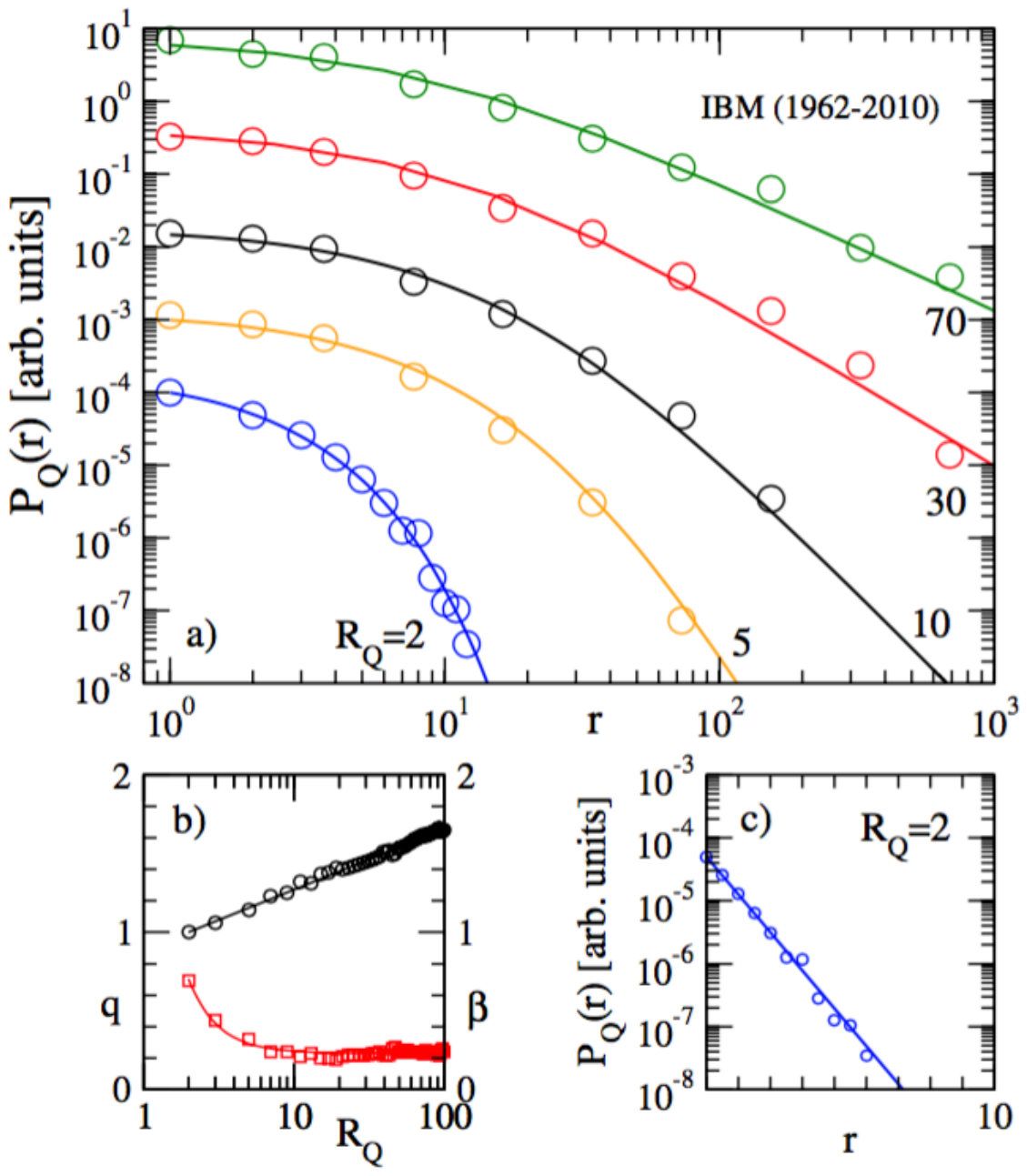}
\end{center}
\vspace{-6.5cm}
\caption{(a) The distribution function of  the inter-occurence times for the daily price returns of IBM during the period 1962-2010.
The full lines show the fitted 
q-exponentials. For $R_Q=2,\,5, \,10,\,30$, and $R_Q=70$ (in units of days), from bottom to top. The full lines show the fitted exponentials.
The dependence of the parameters $\beta$ (squares, lower curve) and $q$ (circles, upper curve) on $R_Q$ in the $q$-exponential is shown in panel (b). Panel (c) confirms that for $R_Q$ = 2 the distribution function is a simple exponential. The straight line is $2^{-r}$. From \cite{LudescherTsallisBunde2011}. Basically the same results are valid for minute up to month price returns \cite{LudescherBunde2014}.
}
\label{fig2_ibm}
\end{figure}

\begin{figure}[h!]
\begin{center}
\includegraphics[width=10cm]{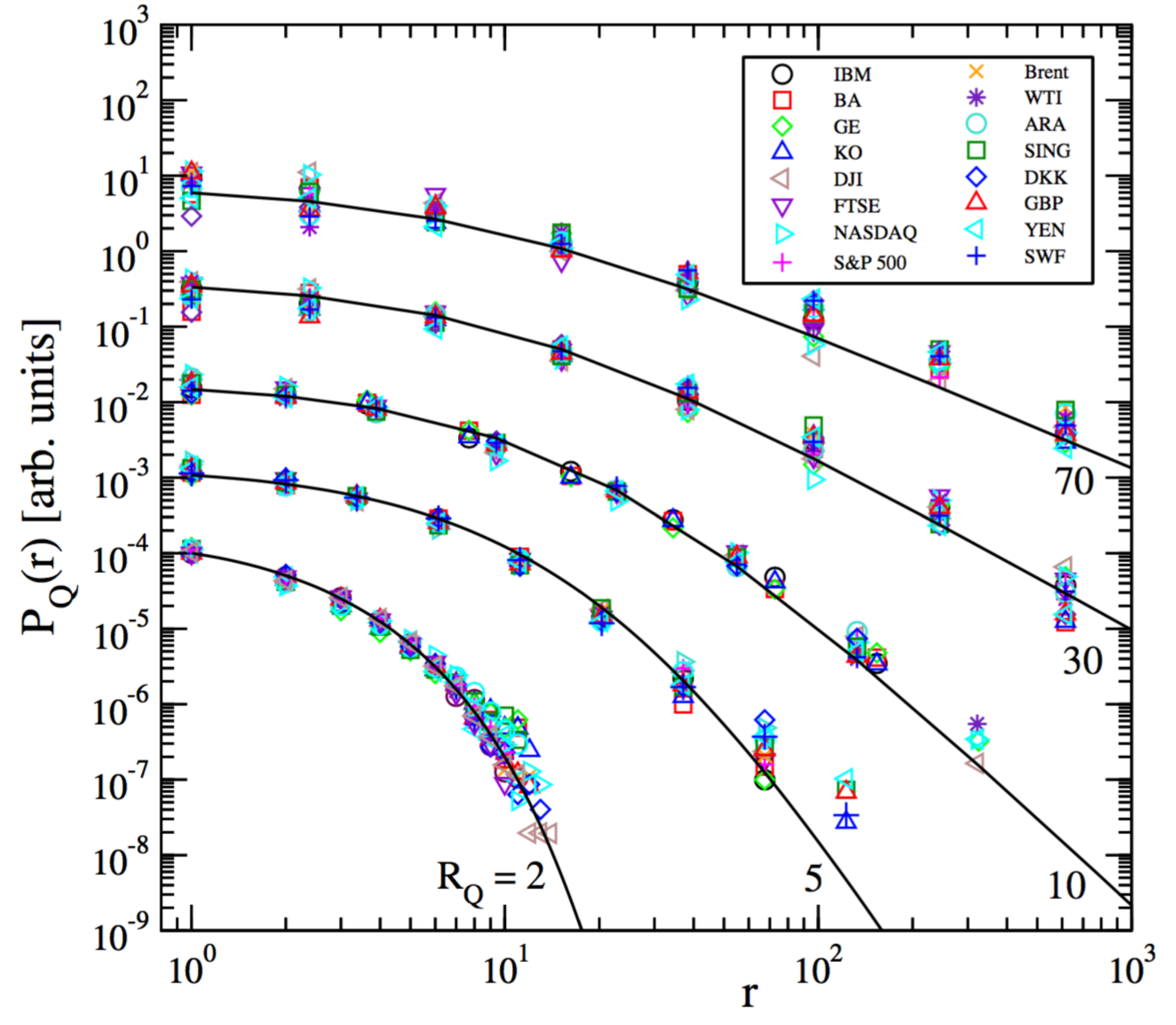}
\end{center}
\caption{The distribution function of the interoccurrence times (as in Fig. \ref{fig2_ibm}) for the daily price returns of 16 examples of financial data, taken from different asset classes (stocks, indices, currencies, commodities). The assets are i) the stocks of IBM, Boeing (BA), General Electric (GE), Coca-Cola (KO), ii) the indices Dow Jones (DJI), Financial Times Stock Exchange 100 (FTSE), NASDAQ, S$\&$P 500, iii) the commodities Brent Crude Oil, West Texas Intermediate (WTI), Amsterdam-Rotterdam-Antwerp gasoline (ARA), Singapore gasoline (SING), and iv) the exchange rates of the following currencies vs. the US Dollar: Danish Crone (DKK), British Pound (GBP), Yen, Swiss Francs (SWF). The full lines show the fitted $q$-exponentials, which are the same as in Fig. \ref{fig2_ibm}. From \cite{LudescherTsallisBunde2011}.
}
\label{fig3_top}
\end{figure}

In Fig. \ref{IBM-Series} we see a typical time series of price returns. The red line at $Q \simeq -0.037$ indicates our choice for checking the inter-occurrence times. The average inter-occurrence time for this value of $Q$ is $R_Q=70$. The value $R_Q$ monotonically increases with $|-Q|$: see Fig. \ref{RQ}. The distributions of inter-occurrence times corresponding to typical values of $R_Q$ are indicated in Fig. \ref{fig2_ibm}. Those corresponding to 16 different financial data are indicated in Fig. \ref{fig3_top}. They are all well fitted through $P_Q(r) \propto e_q^{-\beta\,r}$, where $q=1+q_0\ln(R_Q/2)$ with $q_0 \simeq 0.168$ .  The simplicity of the analytical expression for  $P_Q(r)$ makes easy the calculation of the risk function, which can be shown to be (see \cite{LudescherTsallisBunde2011} for details)
\begin{equation}
W_Q(t;\Delta t)=1-\frac{e_{{\tilde q}}^{-(\beta/{\tilde q})(t+ \Delta t)}}{e_{{\tilde q}}^{-(\beta/{\tilde q})t}}
\end{equation}
with 
\begin{equation}
{\tilde q} = \frac{1}{2 - q} \,. 
\end{equation}

\subsection{Inter-occurrence times in seismic time series}

\begin{figure}[h!]
\vspace{-4.0cm}
\begin{center}
\includegraphics[width=17cm]{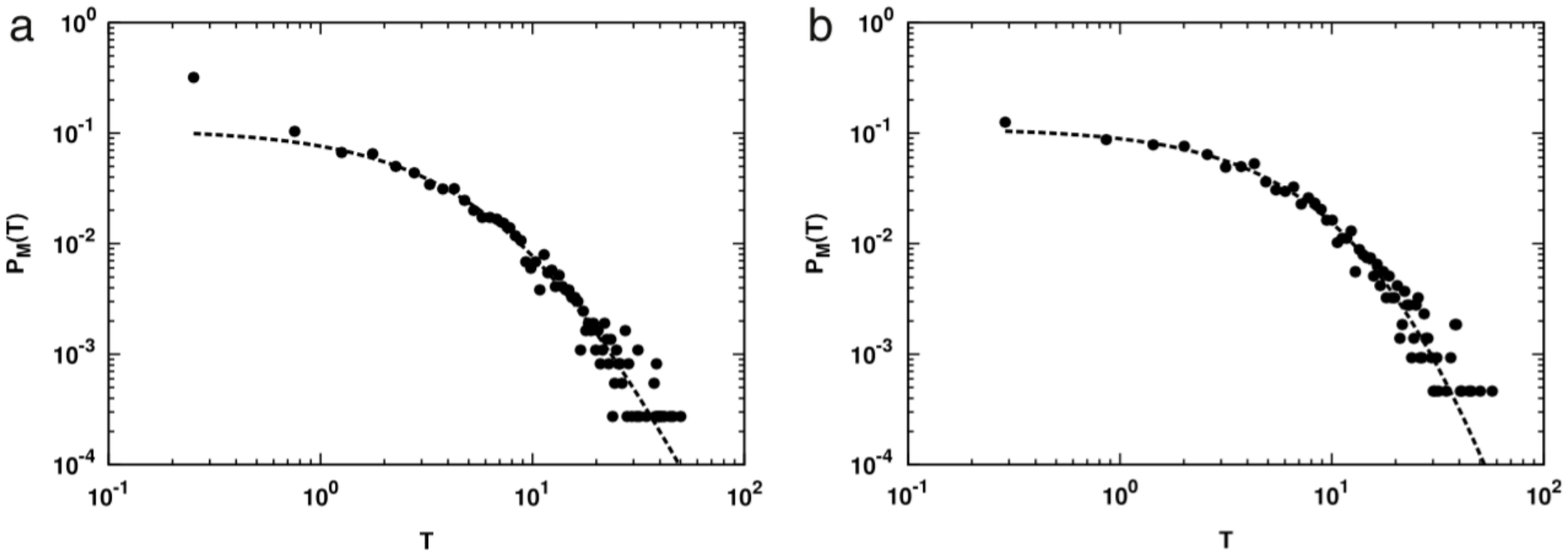}
\end{center}
\vspace{-4.0cm}
\caption{(a) Plot of the inter-event time distribution $P_M (T )$ versus the inter-occurrence time $T$ (in days) for earthquake magnitudes $M \ge 4.1$ for the entire dataset considered (i.e., including the aftershocks). The dashed curve is the fit of the data (in filled circles) through $P_M (T ) \propto e_{q_{T1}}^{- \beta_1 \,T}$ with $q_{T 1}= 1.24 \pm 0.054$ .(b) Same as in (a) but for the corresponding declustered dataset (i.e., excluding the aftershocks). In this case,  $q_{T2} =1.14 \pm 0.057$ . From \cite{AntonopoulosMichasVallianatosBountis2014}.
}
\label{greece}
\end{figure}

Earthquake inter-event times can be analyzed as follows. We choose a threshold value for the magnitude of earthquakes (e.g., 4.1 in the study of seismic activity in Greece during the period 1976-2009: see details in \cite{AntonopoulosMichasVallianatosBountis2014}), and then measure the inter-occurrence times $T$ between earthquakes stronger than the chosen threshold. An illustration is shown in Fig. \ref{greece}.

\subsection{Inter-occurrence distances in genomes}

\begin{figure}[h!]
\vspace{-4.0cm}
\begin{center}
\includegraphics[width=15cm]{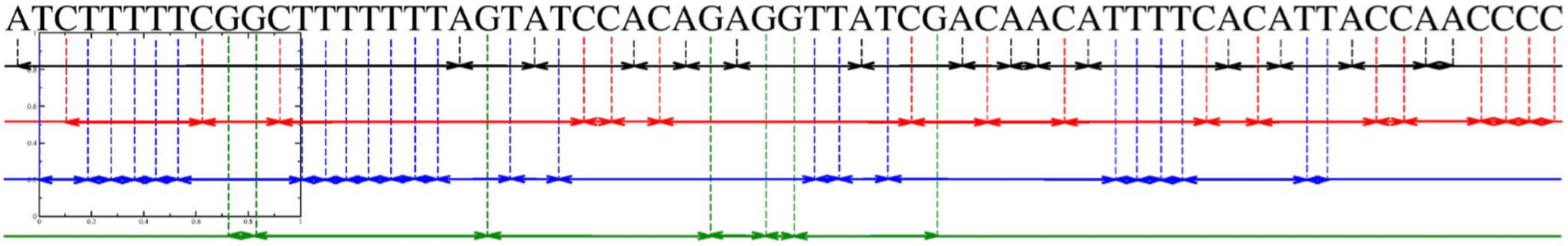}
\end{center}
\vspace{-4.5cm}
\caption{The procedure of the assessment of the four internucleotide interval sequences from the DNA primary sequence. From \cite{BogachevKayumovBunde2014}.
}
\label{DNA}
\end{figure}

\begin{figure}[h!]
\vspace{-6.0cm}
\begin{center}
\includegraphics[width=15cm]{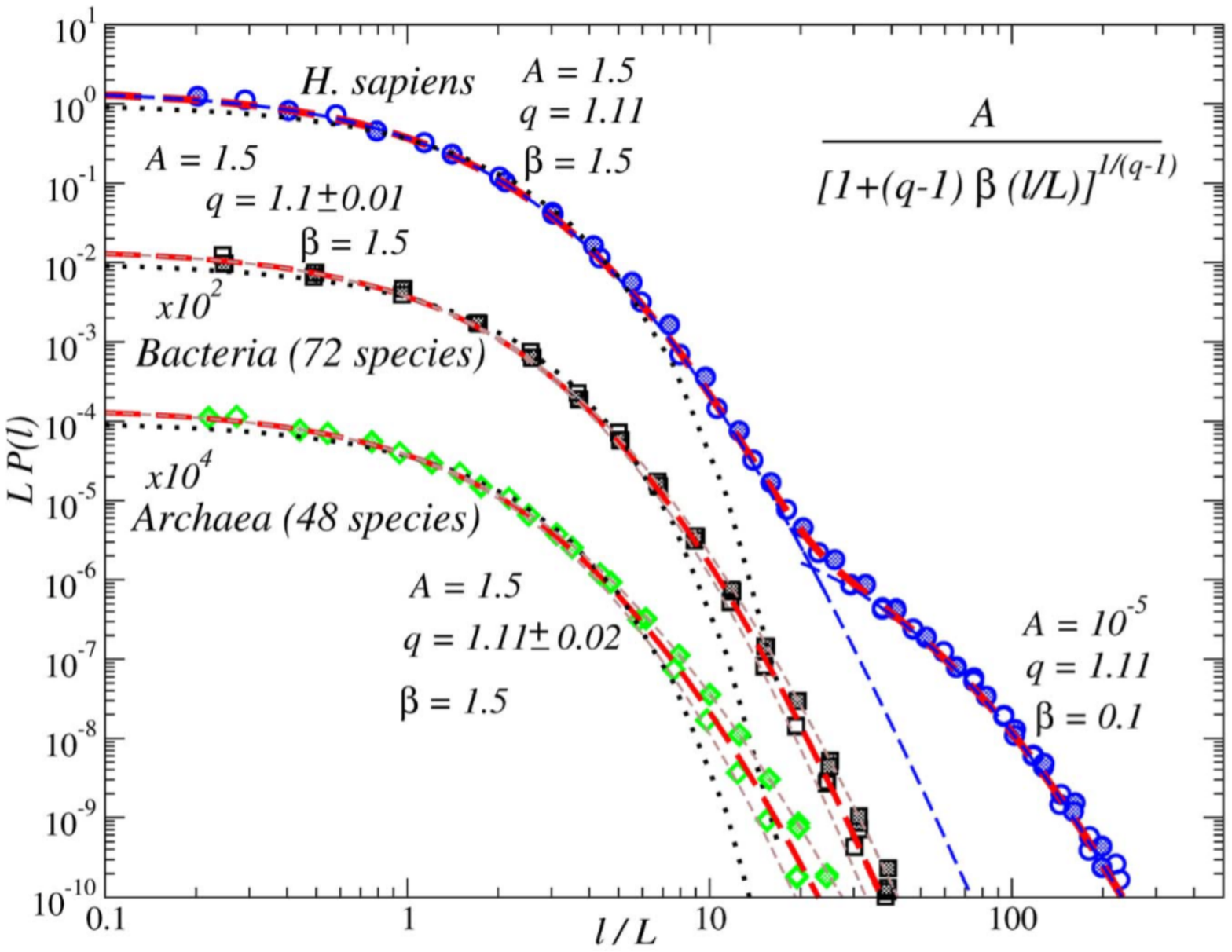}
\end{center}
\vspace{-6.0cm}
\caption{PDFs of the inter-nucleotide intervals A-A, T-T (open symbols); G-G, C-C (full symbols) in the DNA sequences from {\it H. Sapiens} and {\it Bacteria} full genomes (in scaled form). Dashed lines show the best fits by a q-exponential distribution $A/[1+(q-1)\beta (l/L)]^{\frac{1}{q-1}}$. While in {\it Bacteria} the approximation by a single $q$-exponential with $q \simeq 1.1$ and $\beta \simeq 1.5$ is possible, in {\it H. Sapiens} a sum of two $q$-exponentials with $q \simeq 1.11$ and $\beta \simeq 1.5$ and $0.1$ makes the best fit. To avoid overlapping, the PDFs for {\it Bacteria} are shifted downwards by two decades. For comparison, dotted lines show corresponding exponential PDFs. From \cite{BogachevKayumovBunde2014}.
}
\label{DNA2}
\end{figure}

Genomes can be analyzed similarly to financial and earthquake inter-occurrence times, where the {\it distance} between equal nucleotides (say distances between successive adenines, or between successive cytosines) plays the role of 'time'. We follow here \cite{BogachevKayumovBunde2014}. The procedure where inter-occurrence distances play the role of inter-occurrence times is described in Fig. \ref{DNA}. In Fig. \ref{DNA2} we report a quite amazing result: the distributions for both {\it H. Sapiens} and a wide range of {\it Bacteria} have not only the same functional ($q$-exponential) form, but also {\it the same values of $q$ and $\beta$}! For the {\it H. Sapiens} it emerges also a second $q$-exponential with still has the same value of $q$. It would certainly be most interesting to find out whether this double-$q$-exponential form also appears in other organisms, for example in other mammals, or whether it is exclusive of {\it H. Sapiens}.

\subsection{Inter-occurrence times in other systems}

Essentially the same laws that have been reported above for financial, seismic, and genomic phenomena emerge in many analytical, computational, experimental and observational results in natural, artificial and social complex systems. One such example is the Coherent Noise Model (see \cite{CelikogluTirnakliQueiros2010} and references therein), where it is observed a distribution  of returns $r$ of the $q$-Gaussian type $e_{q_r}^{-\beta_r r^2}$ (with $q_r>1$), and a distribution of the avalanches sizes $s$ which behaves asymptotically as $s^{-\tau}$ with $\tau \equiv 1/(q_s -1)$ (with $q_s>1$). It has been analytically established, and numerically verified, the following scaling law \cite{CelikogluTirnakliQueiros2010}:
\begin{equation}
q_r=\frac{\tau +2}{\tau} \,,
\end{equation}
i.e., 
\begin{equation}
\frac{2}{q_r-1}=\frac{1}{q_s-1} \,.
\end{equation}
Similar results are obtained for a different model (restricted diffusion) in \cite{TirnakliJensenTsallis2011,ZandTirnakliJensen2015}.

\section{FINAL REMARKS}

We have discussed here a variety of phenomena of the type that typically exists in financial theory, but which also emerge in many other complex systems such as earthquakes and genetics. More precisely, the distribution of inter-occurrence times (or distances) appears again and again to be of the $q$-exponential form. This universal law, though with values for $(q, \beta)$ that depend on the specific universality class that we are focusing on, plays a simple and relevant role. The calculation of the specific values for these parameters is in principle doable from first principles, meaning from the microscopic dynamics of the system. However, this dynamics is, most frequently, either unknown or mathematically intractable. This is why in many (but by no means all) practical cases, the values for $(q, \beta)$ are determined through fitting. This situation is in fact epistemologically not different from our knowledge of the orbits of the planets. Indeed, their elliptic form is easily established within Newtonian mechanics, however the precise determination of the size and orientation of the ellipses requires the knowledge of the initial conditions of all the masses of the Solar System, which is of course unavailable! Astronomers use therefore the Keplerian elliptic approximation and determine the specific parameters through fitting.

In the present review we have primarily focused on the nonadditive entropy $S_q$, and the scaling laws associated with it. Many other (surely over fifty!) anomalous entropic functionals are by now available in the literature, but their description, even brief, remains out of the present scope. Also remain out of the present aim the various very interesting inter-twinnings with generalized central limit theorems \cite{UmarovTsallisSteinberg2008,UmarovTsallisGellMannSteinberg2010,VignatPlastino2007,Hilhorst2010}.
Finally, the knowledge of the necessary and sufficient conditions for a complex system to exhibit $q$-exponential distributions for inter-occurrence times would be very welcome, but it presently remains as a very challenging open question.  

\section*{Acknowledgments}
F.C. Alcaraz, C. Beck, E.P. Borges, T. Bountis, A. Bunde, L.J.L. Cirto, A. Coniglio, E.M.F. Curado, M. Gell-Mann, P. Grigolini, G. Ruiz, H.J. Herrmann, M. Jauregui, H.J. Jensen, J. Ludescher, F.D. Nobre, A. Pluchino, S.M.D. Queiros, A. Rapisarda, P. Rapcan, P. Tempesta, U. Tirnakli, S. Umarov, and many others, are warmly acknowledged for old and recent fruitful discussions. 
I also acknowledge partial financial support by the Brazilian agencies CNPq and FAPERJ, and by the John Templeton Foundation-USA.


\section*{References}

\begin{thebibliography}{00}

\bibitem{Clausius1865}R. Clausius, 
{\it \"Uber verschiedene f\"ur die Anwendung bequeme Formen der Hauptgleichungen der mechanischen W\"armetheorie}, 
Annalen der Physik {\bf 125}, 353 (1865).

\bibitem{Boltzmann1872}L. Boltzmann, {\it Weitere Studien \"uber das W\"armegleichgewicht unter Gas molek\"ulen} [{\it Further Studies on Thermal Equilibrium Between Gas Molecules}], Wien, Ber. {\bf 66}, 275 (1872).

\bibitem{Boltzmann1877}L. Boltzmann,  {\it \"Uber die Beziehung eines allgemeine mechanischen Satzes zum zweiten Haupsatze der W\"armetheorie}, Sitzungsberichte, K. Akademie der Wissenschaften in Wien, Math.-Naturwissenschaften {\bf 75}, 67-73 (1877); English translation ({\it On the Relation of a General Mechanical Theorem to the Second Law of Thermodynamics}) in S. Brush, {\it Kinetic Theory}, Vol. {\bf 2}: {\it Irreversible Processes}, 188-193 (Pergamon Press, Oxford, 1966).

\bibitem{Gibbs1902}J.W. Gibbs, {\it Elementary Principles in Statistical Mechanics -- Developed with Especial Reference to the Rational Foundation of Thermodynamics} (C. Scribner's Sons, New York, 1902; Yale University Press, New Haven, 1948); OX Bow Press, Woodbridge, Connecticut, 1981) (page 35). 

\bibitem{Tsallis1988}C. Tsallis, J. Stat. Phys. {\bf 52}, 479 (1988) [First appeared as preprint in 1987: CBPF-NF-062/87, ISSN 0029-3865, Centro Brasileiro de Pesquisas Fisicas, Rio de Janeiro].

\bibitem{GellMannTsallis2004}M. Gell-Mann and C. Tsallis, eds.  {\it Nonextensive Entropy - Interdisciplinary Applications},  (Oxford University Press, New York, 2004). Soon after publication, the inadvertence became obvious, namely that {\it Nonextensive Entropy} was a misname: it should have been {\it Nonadditive Entropy} instead.  Indeed, nonadditive entropic functionals are introduced precisely in order to have thermodynamically extensive entropies for systems involving certain classes of strong correlations between their elements.
The point is properly discussed in \cite{Tsallis2004} and also in \cite{TsallisGellMannSato2005a,Tsallis2009}, but the title of the book was unfortunately not corrected in due time.

\bibitem{Tsallis2004}C. Tsallis, pages 1- 53, in {\it Nonextensive Entropy - Interdisciplinary Applications}, M. Gell-Mann and C. Tsallis (Oxford University Press, New York, 2004). 

\bibitem{TsallisGellMannSato2005a}C. Tsallis, M. Gell-Mann and Y. Sato, Proc. Nat. Acad. Sci. {\bf 102}, 15377 (2005).

\bibitem{Tsallis2009}C. Tsallis, {\it Introduction to Nonextensive Statistical Mechanics - Approaching a Complex World} (Springer, New York, 2009).

\bibitem{Tsallis2004Cagliari}C. Tsallis, 
Physica A {\bf 340}, 1
(2004). 

\bibitem{Tsallis2009b}C. Tsallis, 
Eur. Phys. J. A {\bf 40}, 257
(2009).

\bibitem{CuradoTempestaTsallis2016}E.M.F. Curado, P. Tempesta and C. Tsallis, Annals Phys. (2016), in press.

\bibitem{Herrmann2013}H.J. Herrmann, private communication (2013).

\bibitem{Penrose1970}O. Penrose, {\it Foundations of Statistical Mechanics: A Deductive Treatment} (Pergamon, Oxford, 1970), p. 167.

\bibitem{Ubriaco2009}M.R. Ubriaco, 
Phys. Lett. A {\bf 373}, 2516
(2009).

\bibitem{TsallisCirto2013}C. Tsallis and L.J.L. Cirto, 
Eur. Phys. J. C {\bf 73}, 2487 (2013).

\bibitem{RibeiroTsallisNobre2013}M.S. Ribeiro, C. Tsallis and F.D. Nobre, 
Phys. Rev. E {\bf 88}, 052107 (2013).

\bibitem{RibeiroNobreTsallis2014}M.S. Ribeiro, F.D. Nobre and C. Tsallis, 
Phys. Rev. E {\bf 89}, 052135 (2014).

\bibitem{HanelThurner2011a}R. Hanel and S. Thurner, 
Europhys. Lett. {\bf 93}, 20006 (2011).

\bibitem{HanelThurner2011b}R. Hanel and S. Thurner, 
EPL {\bf 96}, 50003 (2011).

\bibitem{HanelThurnerGellMann2014}R. Hanel, S. Thurner and M. Gell-Mann,  
Proc. Nat. Acad. Sci. {\bf 111}, 6905
(2014).

\bibitem{Tempesta2011}P. Tempesta, 
Phys. Rev. E {\bf 84}, 021121 (2011). 

\bibitem{Tempesta2015}P. Tempesta, 
Proc. R. Soc. A {\bf 471}, 20150165 (2015).

\bibitem{CarideTsallisZanette1983}A.O. Caride, C. Tsallis and S. I. Zanette,
Phys. Rev. Lett. {\bf 51}, 145
(1983); {\bf 51}, 616 (1983).

\bibitem{CarusoTsallis2008}F. Caruso and C. Tsallis, Phys. Rev. E {\bf 78}, 021101 (2008).

\bibitem{SaguiaSarandy2010}A. Saguia and M.S. Sarandy, Phys. Lett. A {\bf 374}, 3384 (2010).

\bibitem{Alcaraz1987}F.C. Alcaraz, 
J.  Phys. A: Math. Gen. {\bf 20},  2511
(1987).

\bibitem{Alcaraz1990}F.C. Alcaraz and M.J. Martins, 
J. Phys. A: Math. Gen. {\bf 23}, L1079
(1990).

\bibitem{TsallisHaubold2015}C. Tsallis and H.J. Haubold, 
EPL {\bf 110}, 30005 (2015).

\bibitem{CalabreseCardy2004}P. Calabrese and J. Cardy, J. Stat. Mech.: Theory Exp.  (2004) 
P06002; see also C. Holzhey, F. Larsen and F. Wilczek, Nucl. Phys. B {\bf 424}, 443  (1994).

\bibitem{Mendes1997}R.S. Mendes, Physica A {\bf 242}, 299 (1997).

\bibitem{PlastinoPlastino1997}A. Plastino and A.R. Plastino, Phys. Lett. A {\bf 226}, 257 (1997).

\bibitem{Callen1985}H.B. Callen, {\it Thermodynamics and an Introduction to Thermostatistics}, 2nd Edition (John Wiley, 1985).

\bibitem{JundKimTsallis1995} P. Jund, S.G. Kim and C. Tsallis,
Phys. Rev. B {\bf 52}, 50 (1995).

\bibitem{Grigera1996}J.R. Grigera,
Phys. Lett. A {\bf 217}, 47 (1996).

\bibitem{CannasTamarit1996}S.A. Cannas and F.A. Tamarit,
Phys. Rev. B {\bf 54}, R12661 (1996).

\bibitem{SampaioAlbuquerqueMenezes1997}L.C. Sampaio, M.P. de Albuquerque and F.S. de Menezes,
Phys. Rev. B {\bf 55}, 5611 (1997).

\bibitem{AnteneodoTsallis1998}C. Anteneodo and C. Tsallis, Phys. Rev. Lett. {\bf 80}, 5313 (1998).

\bibitem{CurilefTsallis1999}Curilef and C. Tsallis,
Phys. Lett. A {\bf 264}, 270 (1999).

\bibitem{AndradePinho2005}R.F.S. Andrade and S.T.R. Pinho,
Phys. Rev. E {\bf 71}, 026126 (2005).

\bibitem{BinekPolisettyHeMukherjeeRajeshRedepenning2006}Ch. Binek, S. Polisetty, X. He, T. Mukherjee, R. Rajesh and J. Redepenning, 
Phys. Rev. B {\bf 74}, 054432 (2006).

\bibitem{diffusion}C.A. Condat, J. Rangel and P.W. Lamberti,
Phys. Rev. E {\bf 65}, 026138 (2002).

\bibitem{RegoLucenaSilvaTsallis1999}H.H.A. Rego, L.S. Lucena, L.R. da Silva and C. Tsallis,
Physica A {\bf 266}, 42 (1999).

\bibitem{FulcoSilvaNobreRegoLucena2003}U.L. Fulco, L.R. da Silva, F.D. Nobre, H.H.A. Rego and L.S. Lucena,
Phys. Lett. A {\bf 312}, 331 (2003).

\bibitem{BorlandMenchero1999}L. Borland and J.G. Menchero, 
in {\it Nonextensive Statistical Mechanics and Thermodynamics}, eds. S.R.A. Salinas and C. Tsallis, Braz. J. Phys. {\bf 29}, 169 (1999). 

\bibitem{BorlandMencheroTsallis2000}L. Borland, J.G. Menchero and C. Tsallis, 
Phys. Rev. B {\bf 61}, 1650 (2000).

\bibitem{CampaGiansantiMoroniTsallis2001}A. Campa, A. Giansanti, D. Moroni and C. Tsallis,
Phys. Lett. A  \textbf{286}, 251 (2001).

\bibitem{Bekenstein1973a}
J.D. Bekenstein,
Phys. Rev. D \textbf{7}, 2333 (1973).

\bibitem{Bekenstein1973b}J.D. Bekenstein,
Phys. Rev. D \textbf{9}, 3292 (1974).

\bibitem{Hawking1974a}S.W. Hawking,
Nature  \textbf{248}, 30 (1974).

\bibitem{Hawking1974b}S.W. Hawking,
Phys. Rev. D  \textbf{13}, 191 (1976).

\bibitem{tHooft1985}G. 't Hooft, Nucl. Phys. B {\bf 256}, 727 (1985).

\bibitem{tHooft1990}G. 't Hooft,
Nucl. Phys. B  \textbf{355}, 138 (1990), and references therein.

\bibitem{Susskind1993}L. Susskind,
Phys. Rev. Lett.  \textbf{71}, 2367 (1993).

\bibitem{Maddox1993}J. Maddox,
Nature  \textbf{365}, 103 (1993).

\bibitem{Srednicki1993}M. Srednicki,
Phys. Rev. Lett.  \textbf{71}, 666 (1993).

\bibitem{StromingerVafa1996}A. Strominger and C. Vafa,
Phys. Lett. B  \textbf{379}, 99 (1996).

\bibitem{MaldacenaStrominger1998}J. Maldacena and A. Strominger,
J. High Energy Phys. \textbf{2}, 014 (1998).

\bibitem{DasandShankaranarayanan2006}S. Das and S. Shankaranarayanan,
Phys. Rev. D {\bf 73}, 121701 (R) (2006).

\bibitem{BrusteinEinhornYarom2006}R. Brustein, M.B. Einhorn and A. Yarom,
J. High Energy Phys. \textbf{01}, 098 (2006).

\bibitem{BorstenDahanayakeDuffEbrahimRubens2009}L. Borsten, D. Dahanayake, M.J. Duff, H. Ebrahim and W. Rubens,
Phys. Rep. {\bf 471}, 113 (2009).

\bibitem{Padmanabhan2009}T. Padmanabhan,
arXiv: 0910.0839 (2009).

\bibitem{Casini2009}H. Casini, Phys. Rev. D {\bf 79}, 024015 (2009).

\bibitem{BorstenDahanayakeDuffMarraniRubens2010}L. Borsten, D. Dahanayake, M.J. Duff, A. Marrani and W. Rubens,
Phys. Rev. Lett. {\bf 105}, 100507 (2010).

\bibitem{Corda2011}C. Corda,
J. High Energy Phys. \textbf{08}, 101 (2011).

\bibitem{KolekarPadmanabhan2011}S. Kolekar and T. Padmanabhan, Phys. Rev. D {\bf 83}, 064034 (2011).

\bibitem{Saida2011}H. Saida, Entropy {\bf 13}, 1611 (2011).

\bibitem{generalrelativity}
L.P. Hughston and K.P. Tod,  {\it An Introduction to General Relativity} (Cambridge University Press, Cambridge, 1990).

\bibitem{generalrelativity_2}
E.F. Taylor and J.A. Wheeler,
{\it Exploring Black Holes: Introduction to General Relativity} (Addison Wesley, San Francisco, 2000).

\bibitem{generalrelativity_3}
S. Weinberg, 
{\it Gravitation and Cosmology: Principles and Applications of the General Theory of Relativity} (John Wiley \& Sons, New York, 1972).

\bibitem{EisertCramerPlenio2010}
J. Eisert, M. Cramer and  M.B. Plenio,
Rev. Mod. Phys. \textbf{82}, 277 (2010).

\bibitem{carlip}
M. Banados, C. Teitelboim and J. Zanelli,
Phys. Rev. Lett. {\bf 69}, 1849 (1992).

\bibitem{carlip_2}
S. Carlip,
Class. Quantum Grav. {\bf 12}, 2853 (1995).

\bibitem{carlip_3}
S. Carlip, {\it Quantum Gravity in 2+1 Dimensions} (Cambridge University Press, Cambridge, 1998).

\bibitem{RuizTsallis2012}G. Ruiz and C. Tsallis,
Phys. Lett. A {\bf 376}, 2451 (2012).

\bibitem{Touchette2013}H. Touchette,
Phys. Lett. A {\bf 377}, 436 (2013).

\bibitem{RuizTsallis2013}G. Ruiz and C. Tsallis,
Phys. Lett. A {\bf 377}, 491 (2013).

\bibitem{reminder}Let us remind that the $q$-Gaussians for $q>1$ are also called {\it $\kappa$-distributions} in the plasma physics community, and also occasionally {\it generalized Lorentzians}; for special rational values of $q$ they correspond to the Student's $t$-distributions for an integer number of degrees of freedom. For special rational values of $q<1$, $q$-Gaussians correspond to the $r$-distributions (which have a compact support).

\bibitem{TsallisAnjosBorges2003}C. Tsallis, J.C. Anjos and E.P. Borges,
Phys. Lett. A {\bf 310}, 372 (2003).

\bibitem{TirnakliJensenTsallis2011}U. Tirnakli, H.J. Jensen and C. Tsallis, 
Europhys. Lett. {\bf 96}, 40008 (2011).

\bibitem{ZandTirnakliJensen2015}J. Zand, U. Tirnakli and H.J. Jensen, 
J. Phys. A: Math. Theor. {\bf 48}, 425004 (2015).

\bibitem{PluchinoRapisardaTsallis2007}A. Pluchino, A. Rapisarda and C. Tsallis, 
Europhys. Lett. {\bf 80}, 26002 (2007).

\bibitem{TamaritCannasTsallis1998}F.A. Tamarit, S.A. Cannas and C. Tsallis,
Eur. Phys. J. B {\bf 1}, 545 (1998).

\bibitem{AnteneodoTsallis1997}C. Anteneodo and C. Tsallis,
J. Mol. Liq. {\bf 71}, 255 (1997).

\bibitem{TirnakliTsallisLyra1999}U. Tirnakli, C. Tsallis and M.L. Lyra,
Eur. Phys. J. B {\bf 11}, 309 (1999).

\bibitem{RodriguezSchwammleTsallis2008}A. Rodriguez, V. Schwammle and C. Tsallis, JSTAT P09006 (2008); R.
Hanel, S. Thurner and C. Tsallis, Eur. Phys. J. B {\bf 72}, 263 (2009); A. Rodriguez, C. Tsallis, J. Math. Phys. {\bf 51}, 073301 (2010).

\bibitem{AndradeSilvaMoreiraNobreCurado2012}J. S. Andrade Jr., G.F.T. da Silva, A.A. Moreira, F.D. Nobre and E.M.F. Curado, Phys. Rev. Lett. {\bf 105}, 260601 (2010); Y. Levin and R. Pakter, Phys. Rev. Lett. {\bf 107}, 088901 (2011); J. S. Andrade Jr., G.F.T. da Silva, A.A. Moreira, F.D. Nobre and E.M.F. Curado, Phys. Rev. Lett. {\bf 107}, 088902 (2011); M.S. Ribeiro, F.D. Nobre and E.M.F. Curado,
in Special Issue {\it Tsallis Entropy}, ed. A. Anastasiadis, Entropy  {\bf 13}, 1928 (2011); M.S. Ribeiro, F.D. Nobre and E.M.F. Curado, Phys. Rev. E {\bf 85}, 021146 (2012); M.S. Ribeiro, F.D. Nobre and E.M.F. Curado,
Eur. Phys. J. B {\bf 85}, 399 (2012). 
 
\bibitem{PlastinoPlastino1995}A.R. Plastino and A. Plastino, Physica A {\bf 222}, 347 (1995); C. Tsallis and D.J. Bukman, Phys. Rev. E {\bf 54}, R2197 (1996); L.C. Malacarne, R.S. Mendes, I.T. Pedron and E.K. Lenzi,
Phys. Rev. E {\bf  65}, 052101 (2002); C. Tsallis and Z.G. Arenas, 
EPJ {\bf 71}, 00132 (2014); C. Tsallis and E.K. Lenzi, 
Chemical Physics {\bf 284} (1-2), 341-347 (2002);  O'Malley, V.V. Vesselinov and J.H. Cushman, 
Phys. Rev. E {\bf 91}, 042143 (2015). 

\bibitem{pedron}I.T. Pedron, R.S. Mendes, T.J. Buratta, L.C. Malacarne and E.K. Lenzi, Phys. Rev. E {\bf 72}, 031106 (2005).

\bibitem{AnteneodoTsallis2003}L. Borland, Phys. Lett. A {\bf 245}, 67 (1998); L. Borland, F. Pennini, A. R. Plastino and A. Plastino Eur. Phys. J. B
{\bf 12}, 285 (1999); C. Anteneodo and C. Tsallis, J. Math. Phys. {\bf 44}, 5194 (2003);  V. Schwammle, E.M.F. Curado and F.D. Nobre, Eur. Phys. J. B {\bf 58}, 159
(2007); V. Schwammle, F.D. Nobre and E.M.F. Curado, Phys. Rev. E {\bf 76}, 041123 (2007); M.A. Fuentes and M.O. Caceres, Phys. Lett. A {\bf 372}, 1236 (2008) [to compare with the present paper, it must be done $q \to 2 - q$]; B.C.C. dos Santos and C. Tsallis, Phys. Rev. E {\bf 82}, 061119 (2010); A. Mariz and C. Tsallis, Phys. Lett. A  {\bf 376}, 3088 (2012).

\bibitem{UpadhyayaRieuGlazierSawada2001}A. Upadhyaya, J.-P. Rieu, J.A. Glazier and Y. Sawada, Physica A {\bf 293}, 549 (2001).

\bibitem{ThurnerWickHanelSedivyHuber2003}S. Thurner, N. Wick, R. Hanel, R. Sedivy and L.A. Huber, Physica A {\bf 320}, 475 (2003).

\bibitem{DanielsBeckBodenschatz2004}K.E. Daniels, C. Beck and E. Bodenschatz, Physica D {\bf 193}, 208 (2004).

\bibitem{BurlagaVinas2005}L.F. Burlaga and A.F.Vinas, Physica A {\bf 356}, 375 (2005); L.F. Burlaga, A.F. Vinas, N.F. Ness and M.H. Acuna, Astrophys. J. {\bf 644}, L83 (2006);
L.F. Burlaga and N.F.  Ness,
Astrophys. J. {\bf 703}, 311  (2009);
L.F. Burlaga, N.F. Ness and M.H. Acuna,
Astrophys. J. {\bf 691}, L82 (2009);
J. Cho and A. Lazarian,
Astrophys. J. {\bf 701}, 236 (2009);
L.F. Burlaga and N.F. Ness,
Astrophys. J. {\bf 725}, 1306 (2010) 1306;
A. Esquivel and A. Lazarian,
Astrophys. J. {\bf 710}, 125 (2010);
L.F. Burlaga and N.F. Ness,
Astrophys. J. {\bf 737}, 35   (2011).

\bibitem{DouglasBergaminiRenzoni2006}P. Douglas, S. Bergamini and F. Renzoni, Phys. Rev. Lett. {\bf 96}, 110601 (2006).

\bibitem{LutzRenzoni2013}E. Lutz and F. Renzoni, 
Nature Physics {\bf 9}, 615-619 (2013).

\bibitem{CombeRichefeuStasiakAtman2015}G. Combe, V. Richefeu, M. Stasiak and A.P.F. Atman, 
Phys. Rev. Lett. {\bf 115}, 238301 (2015).

\bibitem{LiuGoree2008}B. Liu and J. Goree, Phys. Rev. Lett. {\bf 100}, 055003 (2008); R.M. Pickup, R. Cywinski, C. Pappas, B. Farago and P. Fouquet,
Phys. Rev. Lett. {\bf 102}, 097202 (2009); R.G. DeVoe, Phys. Rev. Lett. {\bf 102}, 063001 (2009).

\bibitem{CMS}CMS Collaboration, J. High Energy Phys. {\bf 02}, 041 (2010); CMS Collaboration, Phys. Rev. Lett. {\bf 105}, 022002 (2010); CMS Collaboration, J. High Energy Phys. {\bf 09}, 091 (2010); CMS Collaboration, J. High Energy Phys. {\bf 08}, 086 (2011); CMS Collaboration, J. High Energy Phys. {\bf 05}, 064 (2011); ALICE Collaboration, Phys. Lett. B {\bf 693}, 53 (2010); ALICE Collaboration, Eur. Phys. J. C {\bf 71}, 1594 (2011); ALICE Collaboration, Eur. Phys. J. C {\bf 71}, 1655 (2011); ATLAS Collaboration, New J. Physics {\bf 13}, 053033 (2011); PHENIX Collaboration, Phys. Rev. D {\bf 83}, 052004 (2011); PHENIX Collaboration, Phys. Rev. C {\bf 83}, 024909 (2011); PHENIX Collaboration, Phys. Rev. C {\bf 83}, 064903 (2011); PHENIX Collaboration, Phys. Rev. C {\bf 84}, 044902 (2011);
M. Shao, L. Yi, Z.B. Tang, H.F. Chen, C. Li and Z.B. Xu,
J. Phys. G {\bf 37}, 084104 (2010); A. Tawfik,
Nuclear Phys. A {\bf 859}, 63 (2011); C.Y. Wong, G. Wilk, L.J.L. Cirto and C. Tsallis, 
Phys. Rev. D {\bf 91}, 114027 (2015).

\bibitem{TirnakliBeckTsallis2007}U. Tirnakli, C. Beck and C. Tsallis, Phys. Rev. E {\bf 75}, 040106(R) (2007).

\bibitem{TirnakliTsallisBeck2009}U. Tirnakli, C. Tsallis and C. Beck, Phys. Rev. E {\bf 79}, 056209 (2009).

\bibitem{TsallisTirnakli2010}C. Tsallis and U. Tirnakli, J. Phys. Conf. Series {\bf 201}, 012001 (2010).

\bibitem{NobreRegoMonteiroTsallis2011}F.D. Nobre, M.A. Rego-Monteiro and C. Tsallis, Phys. Rev. Lett. {\bf 106}, 140601 (2011).

\bibitem{NobreRegoMonteiroTsallis2012}F.D. Nobre, M.A. Rego-Monteiro and C. Tsallis, EPL {\bf 97}, 41001 (2012).

\bibitem{CostaAlmeidaFariasAndrade2011}R.N. Costa Filho, M.P. Almeida, G.A. Farias and J.S. Andrade, Phys. Rev. A {\bf 84}, 050102(R) (2011).

\bibitem{NobreRegoMonteiro2012}F.D. Nobre, M.A. Rego-Monteiro and C. Tsallis, EPL {\bf 97}, 41001 (2012).

\bibitem{Mazharimousavi2012}S.H. Mazharimousavi, Phys Rev A {\bf 85}, 034102 (2012).

\bibitem{PlastinoTsallis2013}A.R. Plastino and C. Tsallis, J. Math. Phys. {\bf 54}, 041505 (2013).

\bibitem{CostaAlencarSkagerstamAndrade2013}R.N. Costa Filho, G. Alencar, B.S. Skagerstam and J.S. Andrade, EPL {\bf 101}, 10009 (2013).

\bibitem{RegoMonteiroNobre2013a}M.A. Rego-Monteiro and F.D. Nobre, Phys. Rev. A {\bf 88}, 032105 (2013).

\bibitem{ToranzoPlastinoDehesaPlastino2013}I.V. Toranzo, A.R. Plastino, J.S. Dehesa and A. Plastino, Physica A {\bf 392}, 3945 (2013).

\bibitem{RegoMonteiroNobre2013b}M.A. Rego-Monteiro and F.D. Nobre, J. Math. Phys. {\bf 54}, 103302 (2013).

\bibitem{CurilefPlastinoPlastino2013}S. Curilef, A.R. Plastino, A. Plastino, 
Physica A {\bf 392}, 2631
(2013).

\bibitem{PenniniPlastinoPlastino2014}F. Pennini, A.R. Plastino and A. Plastino, Physica A {\bf 403}, 195 (2014)

\bibitem{CostaBorges2014}B.G. Costa and E.P. Borges, J. Math. Phys. {\bf 55}, 062105 (2014).

\bibitem{TsallisPlastinoZheng1997}C. Tsallis, A.R. Plastino and W.-M. Zheng, 
Chaos, Solitons and Fractals {\bf 8}, 885
(1997).

\bibitem{BaldovinRobledo2004_2}
G. Ruiz and C. Tsallis,
Eur. Phys. J. B {\bf 67}, 577 (2009).

\bibitem{BaldovinRobledo2004_3}
F. Baldovin and A. Robledo,
Europhys. Lett. {\bf 60}, 518 (2002).

\bibitem{BaldovinRobledo2004_4}
G. Casati, C. Tsallis and F. Baldovin,
Europhys. Lett. {\bf 72}, 355 (2005).

\bibitem{BaldovinRobledo2004_5}
E.P. Borges, C. Tsallis, G.F.J. Ananos, P.M.C. de Oliveira,
Phys. Rev. Lett. {\bf 89}, 254103 (2002).

\bibitem{BaldovinRobledo2004_6}
G.F.J. Ananos and C. Tsallis,
Phys. Rev. Lett. {\bf 93}, 020601 (2004).

\bibitem{BaldovinRobledo2004_7}
F. Baldovin and A. Robledo,
Phys. Rev. E {\bf 66}, R045104 (2002).

\bibitem{BaldovinRobledo2004_8}
F. Baldovin and A. Robledo,
 Phys. Rev. E {\bf 69}, 045202(R) (2004).

\bibitem{BaldovinRobledo2004_9}
E. Mayoral and A. Robledo,
Phys. Rev. E {\bf 72}, 026209 (2005).

\bibitem{BaldovinRobledo2004_10}
E. Mayoral, A. Robledo,
Physica A {\bf 340}, 219 (2004).

\bibitem{FuentesTsallisSato2011}M.A. Fuentes, Y. Sato and C. Tsallis, 
Phys. Lett. A  {\bf 375},  2988 (2011).

\bibitem{TsallisGellMannSato2005b}C. Tsallis, M. Gell-Mann and Y. Sato, {\it Extensivity and entropy production}, Europhysics News {\bf 36}, 186 (2005) [Europhysics News Special Issue {\it Nonextensive Statistical Mechanics: New Trends, new perspectives}, eds. J.P. Boon and C. Tsallis (November/December 2005].

\bibitem{LuqueLacasaRobledo2012}B. Luque, L. Lacasa and A. Robledo, 
Phys. Lett. A {\bf 376}, 3625 (2012).

\bibitem{AnanosBaldovinTsallis2005}G.F.J. Ananos, F. Baldovin and C. Tsallis, 
Eur. Phys. J. B {\bf 46}, 409
(2005).

\bibitem{LudescherTsallisBunde2011}J. Ludescher, C. Tsallis and A. Bunde, 
Europhys. Lett. {\bf 95},  68002  (2011).

\bibitem{LudescherBunde2014}J. Ludescher and A. Bunde, 
Phys. Rev. {\bf 90}, 062809 (2014).

\bibitem{AntonopoulosMichasVallianatosBountis2014}C.G. Antonopoulos, G. Michas, F. Vallianatos and T. Bountis, 
Physica A {\bf 409}, 71-77 (2014).

\bibitem{BogachevKayumovBunde2014}M.I. Bogachev, A.R. Kayumov and A. Bunde, 
PLoS ONE {\bf 9} (12), e112534 (2014).

\bibitem{CelikogluTirnakliQueiros2010}A. Celikoglu, U. Tirnakli and S.M.D. Queiros, 
Phys. Rev. E {\bf 82}, 021124 (2010).

\bibitem{NobreTsallis1995}F.D. Nobre and C. Tsallis, 
Physica A {\bf 213}, 337 (1995); Erratum: {\bf 216}, 369 (1995).

\bibitem{LucenaSilvaTsallis1995}L.S. Lucena, L.R. da Silva and C. Tsallis, 
Phys. Rev. E {\bf 51} (6), 6247
(1995).

\bibitem{PluchinoRapisarda2007}A. Pluchino and A. Rapisarda, 
SPIE {\bf 2}, 6802-32 (2007). 

\bibitem{PluchinoRapisardaTsallis2008}A. Pluchino, A. Rapisarda and C. Tsallis, 
Physica A {\bf 387}, 3121
(2008).
  
\bibitem{FigueiredoRochaAmato2008}A. Figueiredo, T.M. Rocha Filho and M.A. Amato, 
Europhys. Lett. {\bf 83}, 30011 (2008).

\bibitem{PluchinoRapisardaTsallis2009}A. Pluchino, A. Rapisarda and C. Tsallis, 
Europhys. Lett. {\bf 85}, 60006 (2009).

\bibitem{FigueiredoRochaAmato2009}A. Figueiredo, T.M. Rocha Filho and M.A. Amato, 
Europhys. Lett. {\bf 85}, 60007 (2009).

\bibitem{CirtoAssisTsallis2014}L.J.L. Cirto, V.R.V. Assis and C. Tsallis, 
Physica A {\bf 393},  286-296 (2014).

\bibitem{ChristodoulidiTsallisBountis2014}H. Christodoulidi, C. Tsallis and T. Bountis, 
EPL {\bf 108}, 40006 (2014).

\bibitem{UmarovTsallisSteinberg2008}S. Umarov, C. Tsallis and S. Steinberg, Milan J. Math. {\bf 76}, 307 (2008); for a simplified version, see S.M.D. Queiros and C. Tsallis, AIP Conference Proceedings {\bf 965}, 21 (New York, 2007).

\bibitem{UmarovTsallisGellMannSteinberg2010}S. Umarov, C. Tsallis, M. Gell-Mann and S. Steinberg, J. Math. Phys. {\bf 51}, 033502 (2010).

\bibitem{VignatPlastino2007}C. Vignat and A. Plastino, J. Phys. A {\bf 40}, F969 (2007);
M. Przystalski,
Phys. Lett. A  {\bf 374}, 123 (2009);
M.G. Hahn, X.X. Jiang and S. Umarov, J. Phys. A {\bf 43}, 165208 (2010).

\bibitem{Hilhorst2010}A gap has been detected a few years ago,  by H.J. Hilhorst, J. Stat. Mech., P10023 (2010), in the proofs of \cite{UmarovTsallisSteinberg2008}, namely that the $q$-Fourier transform has not, for $q \ne 1$, an inverse in the simple sense that the standard Fourier transform does have. This point certainly is interesting on its own, even if the thesis of the theorems are consistent with alternative proofs that do not use the $q$-Fourier transform. Very recently it has been shown by
M. Jauregui and C. Tsallis, Phys. Lett. A {\bf 375}, 2085 (2011), M. Jauregui, C. Tsallis and E.M.F. Curado, J. Stat. Mech., P10016 (2011), A. Plastino and M.C. Rocca,
Physica A {\bf 391}, 4740 (2012), and A. Plastino and M.C. Rocca, Milan J. Math. {\bf 80} (1), 243 (2012),
that an {\it unique} inverse of the $q$-Fourier transform does exist if some supplementary information is given (something which is unnecessary for, and only for, $q=1$). The use of these results in order to fill the mentioned gap in the proofs of the theorems \cite{UmarovTsallisSteinberg2008} remains now to be done.

\end{thebibliography}
\end{document}